\documentclass[conference]{IEEEtran}

\usepackage[T1]{fontenc}
\usepackage[utf8]{inputenc}

\usepackage{cite}
\usepackage{amsmath,amssymb,amsfonts}

\usepackage{pifont}
\usepackage{algorithm}
\usepackage{algpseudocode}

\usepackage{graphicx}
\usepackage{textcomp}
\usepackage{xcolor}

\usepackage{textcomp}

\usepackage[kerning=true]{microtype}

\usepackage{multirow}
\usepackage{multicol}

\usepackage{url}

\usepackage{subcaption}

\usepackage{xspace}
\usepackage{bmpsize}
\usepackage{xcolor}
\usepackage{lipsum}
\usepackage[colorlinks=true,urlcolor=black]{hyperref}
\definecolor{mycitecolor}{rgb}{0.5,0,0}
\definecolor{mylinkcolor}{rgb}{0.0, 0.5, 1.0}
\definecolor{myurlcolor}{rgb}{0.0, 0.28, 0.67}
\hypersetup{citecolor=mycitecolor}
\hypersetup{linkcolor=mylinkcolor}
\hypersetup{urlcolor=myurlcolor}

\def\BibTeX{{\rm B\kern-.05em{\sc i\kern-.025em b}\kern-.08em
    T\kern-.1667em\lower.7ex\hbox{E}\kern-.125emX}}

\newcommand{\figurespath}{Figures}

\newcommand{\etal}{\textit{et al.\@\xspace}}

\definecolor{burgundy}{rgb}{0.5, 0.0, 0.13}
\definecolor{cadmiumgreen}{rgb}{0.0, 0.42, 0.24}
\definecolor{burntorange}{rgb}{0.8, 0.33, 0.0}
\definecolor{ballblue}{rgb}{0.13, 0.67, 0.8}
\definecolor{azure(colorwheel)}{rgb}{0.0, 0.5, 1.0}
\definecolor{brightmaroon}{rgb}{0.76, 0.13, 0.28}

\newcommand{\our}{\textsc{BaFFLe}}
\newcommand{\layerone}{\textsc{BaFFLe-S}}
\newcommand{\layertwo}{\textsc{BaFFLe-C}}
\newcommand{\layerthree}{\our}
\newcommand{\auditingclients}{validating clients}

\mathchardef\mhyphen="2D 
\newcommand{\xmark}{\ding{55}}%

\newcommand{\defeq}{=}

\newcommand{\getsr}{\gets_{\scriptscriptstyle\$}} 

\newcommand{\spandots}{\,..\,}
\newcommand{\sspan}[2]{[{#1}\spandots{#2}]} 
\renewcommand\vec{\mathbf}

\newcommand{\NN}{\mathbb{N}}
\newcommand{\ZZ}{\mathbb{Z}}

\newcommand{\size}[1]{\lvert #1 \rvert} 

\newcommand{\server}{S}
\newcommand{\client}{C}
\newcommand{\clients}{\mathcal{C}}

\newcommand{\prob}[1]{\Pr \left[ #1 \right]}

\newcommand{\instances}{\mathcal{X}}

\newcommand{\ACC}{\mathrm{acc}}
\newcommand{\ERR}{\mathrm{err}}
\newcommand{\accuracy}[2]{\ACC_{#1}({#2})} 
\newcommand{\error}[2]{\ERR_{#1}({#2})} 

\newcommand{\bdaccuracy}[2]{{\mathrm{acc}^{bd}}_{#1}({#2})} 

\newcommand{\adv}{\mathcal{A}} 

\newcommand{\classifier}{f}

\newcommand{\distr}{\mathcal{D}} 
\algnewcommand\algorithmicforeach{\textbf{foreach}}
\algdef{S}[FOR]{ForEach}[1]{\algorithmicforeach\ #1\ \algorithmicdo}
\algnewcommand{\OR}{\algorithmicor}

\newcommand{\LOF}{\mathsf{LOF}}

\IEEEoverridecommandlockouts

\begin{document}

\title{BaFFLe: Backdoor Detection via\\ Feedback-based Federated Learning}

\author{
    \IEEEauthorblockN{Sebastien Andreina}
    \IEEEauthorblockA{
    \textit{NEC Labs Europe}\\
    sebastien.andreina@neclab.eu}
    \and
    \IEEEauthorblockN{Giorgia Azzurra Marson}
    \IEEEauthorblockA{
    \textit{NEC Labs Europe}\\
    giorgia.marson@neclab.eu}
    \and
    \IEEEauthorblockN{Helen Möllering\textsuperscript{*}\thanks{\textsuperscript{*}Work done while author was working as an intern at NEC Laboratories Europe.}}
    \IEEEauthorblockA{
    \textit{ENCRYPTO/TU Darmstadt}\\
    moellering@encrypto.cs.tu-darmstadt.de}
    \and
    \IEEEauthorblockN{Ghassan Karame}
    \IEEEauthorblockA{
    \textit{NEC Labs Europe}\\
    ghassan@karame.org}
}
\maketitle

\begin{abstract}
Recent studies have shown that federated learning~(FL) is vulnerable to poisoning attacks that inject a backdoor into the global model.
These attacks are effective even when performed by a single client, and undetectable by most existing defensive techniques.
In this paper, we propose \emph{Backdoor detection via Feedback-based Federated Learning} (\our), a novel defense to secure FL against backdoor attacks.
The core idea behind~\our{} is to leverage data of multiple clients not only for training but also for uncovering model poisoning.
We exploit the availability of diverse datasets at the various clients by incorporating a feedback loop into the FL process, to integrate the views of those clients when deciding whether a given model update is genuine or not.
We show that this powerful construct can achieve very high detection rates against state-of-the-art backdoor attacks, even when relying on straightforward methods to validate the  model. Through empirical evaluation using the CIFAR-10 and FEMNIST datasets, we show that by combining the feedback loop with a method that suspects poisoning attempts by assessing the per-class classification performance of the updated model, \our{} reliably detects state-of-the-art backdoor attacks
with a detection accuracy of~100\% and a false-positive rate below~5\%. Moreover, we show that our solution can detect adaptive attacks aimed at bypassing the defense. 
\end{abstract}


\section{Introduction}

Federated learning (FL) is emerging as a powerful paradigm to collaboratively train a machine-learning model among thousands or even millions of participants~\cite{AISTATS:McMahanMRHA17}.
FL involves users' devices in the computation of the machine learning model; here, a central server orchestrates the training of a shared model among several clients.
Each client trains a model locally, on their device, and then all local models are combined to derive a global model summarizing the contributions of all clients.
This process is repeated several times, and the global model is progressively improved within every round of training.

Compared to centralized learning, where model training is operated fully by the server, federated learning features a significant cost reduction (on the server) in terms of computation due to outsourcing and parallelizing the training process.
Involving the clients also enlarges the training set tremendously, which in turn can lead to smarter and more reliable models~\cite{FLgoogleblog}.

In addition to speeding up training and providing better prediction models, FL also promises strong \emph{privacy} guarantees for the clients, in that the training data never leaves the device.
For a truly private solution, a secure aggregation mechanism is put in place that allows clients to combine local updates in a privacy-preserving manner prior to sharing them with the server~\cite{CCS:BonawitzIKMMPRS17}, as otherwise each local model could leak the training data of the corresponding client~\cite{DBLP:conf/sp/NasrSH19}.
While (some sort of) secure aggregation is necessary for protecting the clients' privacy, it comes with a severe security limitation: hiding individual updates prevents clients from being accountable for their contributions to the global model.
This leaves the door open for malicious clients, who can submit manipulated updates in order to tamper with the training process.

A prominent example highlighting the vulnerability of~FL against malicious participants is \emph{model poisoning}, an attack strategy that aims to inject a backdoor into the global model~\cite{AISTATS:BagdasaryanVHES20,ICML:BhagojiCMC19}.
Here, poisoning is operated by malicious clients through carefully crafted updates, and the resulting global model is backdoored in the sense that it assigns a wrong classification to all inputs with a certain (attacker-chosen) feature, while it behaves normally on inputs lacking such feature.
This attack type is inherently hard to detect, because a backdoored model is designed to exhibit adversarial behavior on inputs which are only known to the attacker.

The literature features various proposals to mitigate poisoning attacks on~FL.
A natural approach to limit the effect of malicious clients is to recast existing techniques, developed within Byzantine fault-tolerant distributed learning~\cite{ICML:MhamdiGR18,ACSAC:ShenTS16}, to the FL setting.
However, these techniques have been proven inappropriate as they crucially rely on the training data being uniformly distributed among participants, which is unrealistic for most FL applications.
Other proposals rely on inspecting the local updates submitted by clients to detect poisoning attempts, which makes them incompatible with secure aggregation.
These existing solutions either have been circumvented by adaptive attacks~\cite{FoolsGold}, neglect the privacy of clients~\cite{ARXIV:abs-2002-00211}, or require considerable and computationally-heavy modification to existing~FL algorithms~\cite{nguyen2021flguard}.

In this paper, we tackle the challenge of designing a solution against model poisoning which is fully compliant with currently deployed FL algorithms---including secure aggregation---while being able to resist adaptive adversaries.
At the core of our solution, dubbed \our{} (\emph{Backdoor detection via Feedback-based Federated Learning}), we exploit a peculiar feature of~FL that can effectively help in defeating backdoor attacks. Namely, we observe that in FL settings, multiple private datasets are available which collectively provide a rich and diversified set of labelled data.

To leverage this observation, we design \our{} with the aim to rely on clients not only for training but \emph{also for validating} the global model.
\our{} uses a set of validating clients, refreshed in each training round, to determine if the (global) model-update derived in that round has been subject to a poisoning injection, discarding the update in this case.
That is, clients validate the global model on their local data, and vote for accepting or rejecting the model through a \emph{feedback loop}.
More importantly, the verdict is determined exclusively based on the global model, rather than on individual updates, to ensure compatibility with secure aggregation.
Similarly to the idea of federated training, the core intuition behind our defense is to leverage the multitude of clients participating in the FL process to improve a task otherwise sub-optimally carried out by a single entity (e.g., the coordinating server).

Notice that this approach shares some of the challenges of executing a distributed protocol in a Byzantine setting: malicious clients may misreport their assessment of the model, to have poisoned models declared as ``clean'' (to avoid detection) and clean models deemed as ``poisoned'' (DoS attack). Nevertheless, we show that a proper use of the clients' judgement in the feedback loop can be very powerful in detecting 
backdoor attacks \emph{even when relying on straightforward methods to assess the classification performance.}
Here, we use an off-the shelf method to locally compare (at each client) the classification performance of the updated model with that of the previous model, discarding updates that present unexpected behavior.
To identify ``unexpected behavior'', we analyze the wrong predictions made by the models on a \emph{per-class} basis, and raise a warning whenever the variations of corresponding error rates exceed an empirically determined threshold.

Such a cross-round analysis relies on the intuition that a backdoored model is likely to alter, as a side effect, the misclassification distribution of the model on one or a few classes, \emph{unless the backdoor is optimized for preserving the classification performance on the given validation dataset.}
Crucially, since data is not distributed homogeneously in realistic applications of FL, It is difficult for an adaptive attacker to predict the behavior of the backdoor on the validation sets held by honest clients, and hence to bypass the defense.

We evaluated the effectiveness of our proposals, under various configurations, against the semantic-backdoor attack of Bagdasaryan~\etal~\cite{AISTATS:BagdasaryanVHES20}, in the context of image classification on CIFAR-10 and FEMNIST datasets. Our results show that~\our{} can achieve an accuracy of up to 100\% (and false-positive rate below~5\%), even when the validation method is instantiated with a straightforward misclassification analysis and despite the validation sets at the clients being relatively small.
More importantly, we set our experiments so that none of the validating clients hold backdoor data, thereby considering the worst-case scenario from the defender's perspective.
Our results show that~\our{}---even in this case---can reliably detect poisoning attempts, confirming the intuition that a backdoor injection causes visible side-effect on a dataset for which the attacker did not optimize for.
We also analyzed the detection accuracy against \emph{adaptive attacks} that create poisoned updates such that they bypass our misclassification method on the attacker's validation data.
When evaluated against this stronger attack, which is fully aware of the defense, \our{} still achieves high detection accuracy, confirming the intuition that decentralizing data is an asset by itself in FL.

The remainder of our paper is organized as follows.
In Section~\ref{sec:background}, we provide relevant background and notations about machine learning and federated learning.
In Section~\ref{sec:model}, we specify a security model for backdoor attacks on FL.
We present our proposal to defend against backdoor attacks on FL in Section~\ref{sec:solution},
and we report on our experimental evaluation of~\our{} in Section~\ref{sec:evaluation}.
In Section~\ref{sec:related:work}, we overview closely related work.
We conclude the paper in Section~\ref{sec:conclusion}.

\section{Background}
\label{sec:background}

In this section, we introduce the notations used in the paper, present relevant concepts and terminology, and describe a model-poisoning strategy devised for the FL scenario.
For $a,b \in\NN$ we write $\sspan{a}{b} \defeq \{ x\in\NN : a \leq x \leq b \}$.
Let~$X$ be a (finite) set, and let~$\distr\colon X \to [0,1]$ be a probability distribution.
We denote by $x \gets_\distr X$ the random sampling of an element~$x$ according to distribution~$\distr$;
we write~$x \getsr X$ for the special case of sampling~$x$ uniformly at random from~$X$.

\subsection{Machine Learning}
\label{sec:preliminaries:ML}

A machine-learning classification problem consists of deriving an unknown mapping from an input space~$\instances$ to a set~$Y$ of classes (or labels), given a set of labeled pairs $D = \{(x,y) : x\in X \}$ for some~$X \subseteq \instances$.
Typically, $\instances$ is a vector space, $Y$ is a finite discrete-value set, and the sought mapping between $\instances$ and $Y$ is called \emph{ground truth}.
A \emph{classification model} (a.k.a.~\emph{model} for short) is a mapping $f\colon \instances \to Y$ aimed at emulating the ground truth. 
In supervised learning, the process of deriving a model~$f$ from a labelled dataset~$D$ is called \emph{training}, and~$D$ is called \emph{training set}.
A model should make correct predictions on unseen data, ideally reproducing the ground truth.
The quality of a model~$f$ is measured by its \emph{accuracy} on naturally occurring inputs, defined as
$\accuracy{\distr}{f} \defeq \prob{ f(x) = y : x\gets_\distr \instances}$,
where~$\distr$ denotes the ``natural distribution'' of input samples.
A more pragmatic metric to evaluate the model's classification performance is the \emph{empirical accuracy}.
For a labelled dataset~$D$, we denote the empirical accuracy of~$f$ on \emph{test set}~$D$ by~$\accuracy{D}{\classifier} \defeq \size{ \{(x,y) \in D : \classifier(x) = y\}}/\size{D}$.
\emph{Error} and \emph{empirical error} are defined analogously by considering the wrong predictions made by the model, with  $\error{D}{f} = 1 - \accuracy{D}{f}$.

\subsection{Federated Learning}
\label{sec:preliminaries:FL}

Federated learning provides a decentralized method to train a model in a distributed fashion, leveraging data and resources of several users~\cite{AISTATS:McMahanMRHA17}.
FL is an iterative process, orchestrated by a central server~$\server$, enabling multiple clients to collaboratively train a shared model. In each round, clients perform model-training \emph{locally}, so that their data never have to leave the device, and then aggregate the locally-computed updates to derive an improved version of the shared model.

Throughout the paper, we denote by~$R$ the total number of rounds, by~$\clients$ the set of participating clients, and $N = \size{\clients}$.
Starting from a server-initialized global model~$G^0$, each round~$r\in\sspan{1}{R}$ comprises the following steps:
$\server$ randomly selects~$n \ll N$ clients $\client_1,\dots,\client_n \getsr \clients$ for that round and provides them with  the current model~$G \defeq G^{r-1}$;
each client~$\client_i$ is expected to locally train~$G$ on their private data~$D_i$ to derive an improved local model~$L_i$, and to share the corresponding update~$U_i = L_i - G$ with the server.
Finally, $S$ averages the clients' contributions~$U_1,\dots,U_n$ and integrates them to the global model~$G$, deriving an updated model as 
$G' = G + \frac{\lambda}{N}\sum_{i=1}^{n} \left( L_i - G \right)$,
where~$\lambda$ is the \emph{global learning rate}, controlling the ``fraction'' of the global model which is updated every round; for instance, setting~$\lambda = N/n$ causes~$G$ to be fully replaced by the average of the local models.

Since the local models~$L_i$ (and hence the updates~$U_i$) could leak information about the training data~\cite{DBLP:conf/sp/MelisSCS19,DBLP:conf/sp/NasrSH19}, the aggregation phase can be optionally executed in a privacy-preserving manner, through an interactive protocol called~\emph{secure aggregation}, to ensure that no information about the clients' data is leaked to the server~\cite{CCS:BonawitzIKMMPRS17}.
Upon completion of round~$r$, the current global model is set to~$G^r\defeq G'$.

\section{Security Model}
\label{sec:model}

In this section, we specify a security model for our system.
We discuss backdoor attack generically, their application to FL, and specify a system model and assumptions for our system.

\subsection{Backdoor attacks}

A backdoor attack on a classification model is a type of poisoning attack which injects a backdoor into the model.
Informally, a classification model exhibits \emph{backdoor} behavior if it assigns a wrong label to certain attacker-chosen inputs, called \emph{backdoor instances}, while it behaves normally on all other inputs.
We formalize this concept by defining precise adversarial objectives. Our naming conventions are inspired by the formalism of Chen~\etal~\cite{CORR:abs-1712-05526}.
We say that a \emph{backdoor adversary} $\adv$ is associated with a target label~$y_t \in Y$ and a set~$\instances^* \subset \instances$ of backdoor instances.
Given a classification model~$f\colon \instances \to Y$, $\adv$'s objective is to make~$f$ predict backdoor instances as belonging to the target class, i.e., $f(x) = y_t$ for all~$x\in \instances^*$, while assigning the correct prediction to all other instances, so that the backdoor goes unnoticed.
A typical metric for the attacker's success rate is the \emph{backdoor accuracy} on a set~$X^* \subset \instances^*$ of backdoor instances, defined as the portion of samples in~$X^*$ which are assigned label~$y_t$ by the classifier:
\begin{equation}
\label{eq:backdoor:accuracy}
\bdaccuracy{X^*,y_t}{\classifier}
\defeq \frac{\size{\{ x \in X^* : \classifier(x) = y_t\}}}{\size{X^*}}
\end{equation}
Note that defenders cannot measure~$f$'s backdoor accuracy to detect attacks, as only~$\adv$ knows the backdoor set~$\instances^*$.
There exist various strategies to backdoor a model---all of them relying, in a way or another, on poisoning part of the training data with labelled backdoor samples.

\subsection{Backdoor Attacks in Federated Learning}
\label{sec:background:backdoors:bagdasaryan}

Due to its inherently permissionless design, FL is particularly vulnerable to poisoning attacks:
any edge-device can contribute to the training process, and nothing prevents malicious clients from presenting poisoned updates.
A prominent attack of this type is~\emph{model replacement}~\cite{AISTATS:BagdasaryanVHES20,ICML:BhagojiCMC19}, where the attacker ``replaces'' the global model with a poisoned one, and the backdoor is injected by training a local model using a blend of poisoned data~$(x,y_t)$ with~$x\in X^*$ and correctly labelled data.
This approach is based on the principle of multi-task learning: backdoor data train the model on the adversarial subtask while genuine data ensure good performance on the main task.
Model replacement can be devastating even when just one malicious client submits a poisoned update in a single round of training (single-shot attack)~\cite{AISTATS:BagdasaryanVHES20}.
We will use this attack strategy as a benchmark for evaluating our defense.

\subsection{System model and assumptions}

We aim at devising a defensive strategy for FL to detect whether the global model has been backdoored in a given round.
Notice that in FL, preventing malicious clients from submitting poisoned updates is impossible without inspecting individual updates, which precludes compliance with secure aggregation.
As we seek a defense which does not undermine the privacy of clients' data and is fully compatible with the standard FL algorithms, here we set a milder security objective and require that successful backdoor injections be \emph{detected}---so that the honest participants can react to the poisoning attempt.

We envision our defense to be operated in the following adversarial model.
The attacker controls a number of clients participating in the FL process, and these clients may deviate arbitrarily from the FL protocol (i.e., they are Byzantine).
We assume an honest majority of clients in each training iteration: among the~$n$ chosen clients, up to~$n_M$ can be malicious, where~$n_M < \frac{n}{2}$. We discuss how to estimate~$n_M$ in Section~\ref{sec:evaluation}.
In each round~$r$, the attacker may coordinate its clients towards achieving its goal of backdooring the global model.

\section{Overview \& Design of the Feedback Loop}
\label{sec:solution}

In this section, we present our proposal, Backdoor detection via Feedback-based Federated Learning (\our), and discuss the design of the feedback loop.

\subsection{Intuition and overview}
\label{sec:solution:intuition}

\begin{figure}
	\centering
	\includegraphics[width=1.0\linewidth]{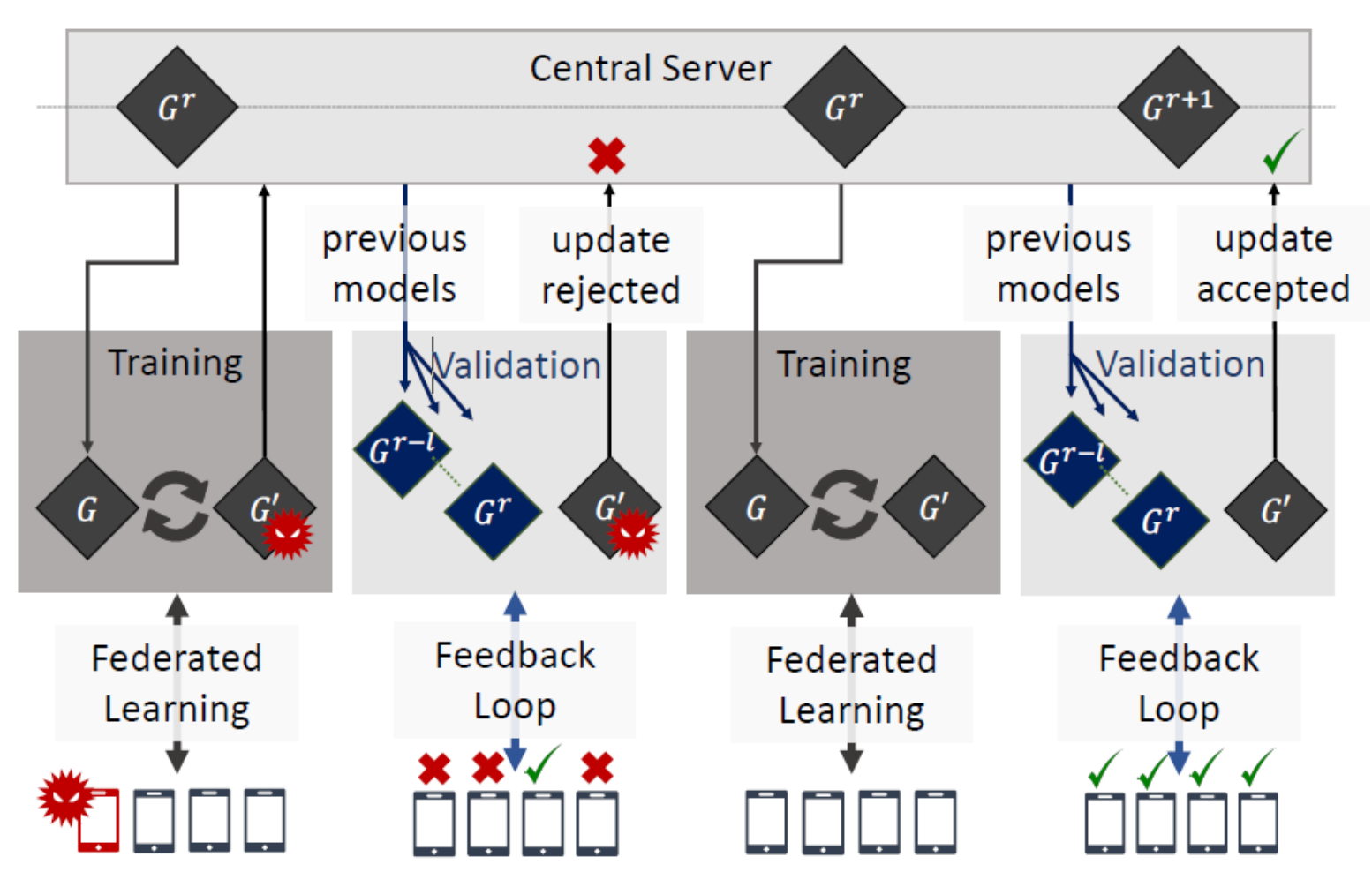}
	\caption{High-level design of~\our{} (two rounds). The first update is detected as poisoned by the validating clients, hence the central server keeps the global model unchanged: $G^r = G$. The second update is accepted and the server sets~$G^{r+1} = G'$.}
	\label{fig:baffle:overview}
\end{figure}

Protecting ML from backdoor attacks is a challenging, if not impossible, task as evidenced by the long sequence of attacks on neural networks and associated defensive strategies (cf.~Section~\ref{sec:related:work}). This difficulty stems from the asymmetry between what the attacker knows about the machine-learning system, and what the defender knows about the attacker. Namely, the backdoor---which is only known to the attacker---can be explicitly chosen to elude detection.

To overcome this challenge, we aim to break this asymmetry between the attacker's and the defender's knowledge by identifying a specific \emph{feature of FL} that can help defending against backdoor attacks:
\emph{the availability of private datasets at the clients which collectively provide a rich and diversified set of labelled data.}

\vspace{1mm}\noindent\textbf{Collaborative model validation.}
In FL, \emph{data is an asset by itself}, and could be leveraged to enhance security.
To this end, we let the clients search for a backdoor in the global model \emph{using their own, private validation data}.
Namely, we propose adding a validation phase to the FL pipeline, so that in each iteration a set of clients, chosen uniformly at random, validates the global model obtained in the previous round.
As clients provide ``feedback'' about the global model, we refer to this process by a \emph{Feedback Loop}. 
Due to the diversity of the clients' data, one immediate effect of having many clients validating the model is to implicitly enlarge and enrich the validation set, thereby increasing the chances of detecting misbehavior.

Letting clients vote on the global model however does introduce a new challenge:
malicious clients can mount denial-of-service attacks by declaring a clean model to be suspicious, or increase the attacker's stealthiness by marking a poisoned model as clean.
Addressing this challenge requires careful calibration of the clients' voting power.

\vspace{1mm}\noindent\textbf{Data unpredictability against adaptive attacks.}
We further look for ubiquitous properties of a backdoor attack on FL that could be turned into a weakness.
Recall that a backdoored model is created by training the original model to perform well on an adversarially-chosen subtask (the \emph{backdoor task}), and for semantic backdoors, training data for the subtask may overlap with training data for the original task.
While typical backdooring strategies can be optimized to preserve the model's classification performance on ordinary validation data, to bypass anomaly-detection methods that rely on measuring model accuracy~\cite{ICML:BhagojiCMC19}, these optimizations assume the attacker has access to a representative set of validation data.
We argue, however, that this assumption is not realistic in a federated setting as validation data held by clients is potentially very diverse.
In particular, in FL there is \emph{substantial data which the attacker does not know/control}, which makes it harder for an attacker to optimize for avoiding detection.

We designed~\our{} based on these observations, which constitute the pillars for the security of our proposal. We present the high-level design of~\our{} in Figure~\ref{fig:baffle:overview}.

\subsection{Design of the Feedback Loop}
\label{sec:solution:feedbackloop}

The feedback loop is a natural augmentation of the FL process, in that it complements federated training with a ``federated validation'' phase.
We now describe the feedback loop in greater detail.

\vspace{1mm}\noindent\textbf{Bootstrapping trust across rounds.}
The feedback loop inspects the global model while training is ongoing, looking for signs of poisoning in every training round.
By design of the FL algorithm, we expect the global model to present \emph{an incremental improvement} in accuracy over the previous model in each round (except for few initial iterations, when the model improves very quickly).
Therefore, we devise a validation method to decide whether the prediction behavior of the global model is ``suspicious''.
We derive relevant information from the prediction behavior of the most recent global models, on a fixed test set, and determine a rejection threshold based on such observations.
This threshold empirically quantifies the expected ``prediction gap'' of a genuine model, based on the predictions of previous models: if the current model exceeds the estimated gap, it is deemed as poisoned and discarded.
Intuitively, we consider early models as a reference for the behavior of ``clean'' models, and bootstrap the trust in these early models towards the subsequent ones, round by round.

\vspace{1mm}\noindent\textbf{Relaxing the trust in early models.} Notice that at the early stages of training, when the global updates evolve sharply from round to round, the model is so immature that the effect of early poisoning is rather negligible, i.e., the backdoor is short-lived, disappearing within only one training round~\cite{ICML:BhagojiCMC19,AISTATS:BagdasaryanVHES20}.
Hence, we do not expect that poisonous injections made during this early phase will affect the effectiveness of~\our{} in the long run---as the model stabilizes---because early injected backdoors tend to vanish immediately.
We validate this behavior experimentally in Section~\ref{sec:evaluation}. Our results show that \our{} is effective even if it is only started as the global model matures, and even if there were poisoning attempts before \our{} has started.

The core operations of the feedback loop are specified in Algorithm~\ref{algo:feedback:loop} and detailed below.

\begin{algorithm}
	\caption{Feedback Loop (for round $r$)}
	\label{algo:feedback:loop}
	\begin{algorithmic}[1]
\small
		    \Statex \textbf{Inputs}
			\Statex Server $\server$: $\{G^t\}_{t=0,\dots,r}, param$
			\Statex Client $C'_i$: $D_i$
			\vspace{2mm}
			\Statex \textbf{Protocol}
			\Statex Server $S$:	
			\State  $(n, \ell, q) \gets param$
			\State  $G \gets G^r$
			\State{Select $\{\mathcal{G}^t\}_{t=0,\dots,\ell}$ from $\{G^t\}_{t=0,\dots,r-1}$}
			\State $history \gets (\mathcal{G}^{0},\dots,\mathcal{G}^{\ell})$
			\State  $C'_1,\dots,C'_n \getsr \clients$
			\For{$i \in \sspan{1}{n}$}
			\State  Send $G$ and $history$ to $C'_i$
			\label{line:feedbackloop:model:history}
			\EndFor

			\Statex \vspace{2mm}Client $C'_i$:
			\State $d_i \gets \textproc{Validate}(G,history,D_i)$
			\State Send $d_i$ to $S$

			\Statex \vspace{2mm}Server $S$:	
			\State Upon receiving~$n$ verdicts:
			\If{$\size{ \{d_i = 1 : i = 1,\dots, n\}} \geq q$}
			\State Reject
			\Else
			\State Accept
			\EndIf
	\end{algorithmic}
\end{algorithm}

\begin{figure}[b]
	\includegraphics[scale=0.6]{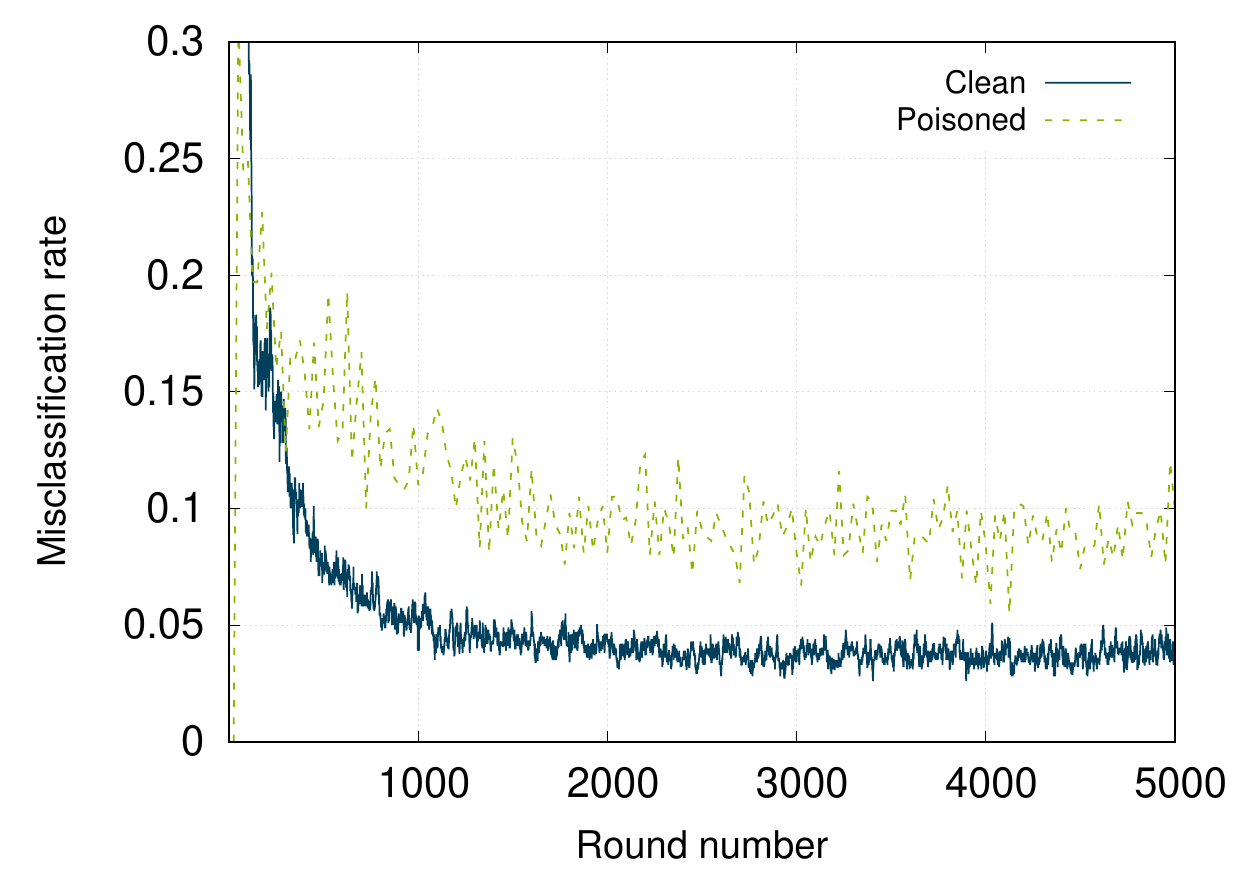}
	\caption{Prediction behavior of clean models \textit{vs} poisoned models on CIFAR-10: per-class error rate w.r.t.~class~7.}
	\label{fig:intuition:wrong-prediction-gap}
	\vspace{-1.5 em}
\end{figure}

\vspace{1mm}\noindent\textbf{Feedback loop specification.}
In the rest of the paper, for each round we refer to the clients that take part in the training process as ``contributors'', and we name ``validating clients'' those which evaluate the updated model and look for signs of poisoning.
Specifically, in every round~$r \in \sspan{1}{R}$, we let the server~$S$ select a set of~$n$ clients, denoted by~$C'_1,\dots,C'_n$, for validating the global model~$G^{r}$ resulting from the updates proposed by the contributors.
Validating clients are responsible for inspecting~$G^{r}$ and reporting to the server whether the model exhibits suspicious behavior.
Each client~$C'_i$ inspects the global model individually, on their own dataset~$D_i$.
For the sake of generality, here we assume a generic validation routine, denoted by~$\textproc{Validate}$, which takes as input
the current global model~$G$,
a history of previously accepted models,
and the client's validation data~$D_i$,
and it returns a verdict~$d_i \in \{0,1\}$ for~$G$, indicating ``clean'' or ``poisoned'' based on the previous models.
We provide one possible instantiation of the validation function in Section~\ref{sec:solution:validation:method}.
Upon receiving all clients' verdicts,%
\footnote{Since some validating clients may not be responding, we may relax this requirement and let the server accept the model by default unless~$q$ many clients suggest rejection.}
the server accepts model~$G^{r}$ as long as sufficiently many clients proposed to accept, else it rejects the global update and starts a new round with ``current'' model~$G^{r+1} \gets G^{r-1}$ (i.e., $G^r$ is discarded).
In what follows, we denote by~$q \in \sspan{1}{n}$ the minimum number of negative verdicts the server requests in order to reject the model, and name it the \emph{quorum threshold}.

\vspace{1mm}\noindent\textbf{Handling malicious votes.}
Having the global model evaluated by clients comes at the risk of giving ``voting power'' to malicious clients, who may deliberately lie, pretending they found indication of poisoning.
Determining an appropriate value for the quorum threshold~$q$ is therefore crucial for the effectiveness of the feedback loop, and care should be taken when weighing the clients' votes.
Clearly, the choice of~$q$ depends on the number of malicious clients that we can tolerate, to ensure that compromised clients can neither cause rejection of genuine models ($n_M < q$), nor outvote honest clients that reject a poisoned model ($n_M \leq n-q+1$), yielding $n_M < q \leq n- n_M$.
The latter condition holds in the ideal scenario where all validating clients have enough data to reliably assess the global model.
Due to the non-uniform distribution of data, a fraction of (honest) clients (i.e., $n-n_M$) $\rho$ might unintentionally provide erroneous judgment of the global model. We argue that this is a peculiarity present in all FL deployments and confirm this observation experimentally in Section~\ref{sec:evaluation}. In fact, we believe that any FL defense that attempts to cluster updates/opinions from clients (whether based on aggregated or separate models) cannot work based on simple majority (i.e., $\rho=0$) as this would give malicious clients a considerable advantage in real FL deployments featuring non-uniform data distributions.
Therefore, we obtain $n_M + (1-\rho)\cdot (n-n_M) < q \leq n - (n_M +(1-\rho)\cdot (n-n_M)) = \rho\cdot (n-n_M)$.
To summarize, setting the quorum threshold to $q := \rho\cdot (n-n_M)$ ensures that aware clients can rightfully reject a poisoned model (true positives), and also ensures that wrong (i.e., naive or malicious) clients cannot reject a honest model (false positives).
By empirically deriving~$\rho$ in our experiments, we can estimate reasonable values for~$q$ and $n_M$ clients.

\section{Model Validation}\label{sec:solution:validation:method}

In this section, we propose an instantiation of the model-validation procedure to detect poisoning attempts. We consider a straightforward approach to assess the classification performance of the model.
Our choice is tailored to the detection of~\emph{semantic backdoors}.
We conjecture that other, dedicated instantiations, tuned to detect different backdoor types than semantic backdoors, can be effective in detecting different backdoor types, too.

\vspace{1mm}\noindent\textbf{Intuition.}
We propose a straightforward validation method to comparing the classification performance of the current model~$G$ with that of recent, previously accepted models~$\mathcal{G}^0,\dots,\mathcal{G}^\ell$ (considered as ``clean'').
Our instantiation relies on observing the model's prediction behavior in a class-wise manner.
Intuitively, we expect that honest updates do not affect significantly the per-class error rates of the global model across subsequent rounds. In contrast, \emph{a freshly injected backdoor is likely to boost the error rate for one or a few classes}.
We validated this expectation experimentally, running FL for an image-classification task on CIFAR-10, and comparing per-class error rates of genuine models and backdoored models obtained via model replacement~\cite{AISTATS:BagdasaryanVHES20}.
Figure~\ref{fig:intuition:wrong-prediction-gap} shows an example of our empirical study.

Based on this observation, we aim at detecting poisoning attempts by monitoring the \emph{variations} in error rates made by the most recently accepted model~$f$ and the updated version~$f'$,  for each label~$y$ and on the same dataset~$D$.
We refer to source-focused error, denoted by~$\error{D}{f}^{y\to \text{\xmark}}$, as the fraction of samples in~$D$ which belong to class~$y$ and are misclassified by~$f$. Similarly, the target-focused error~$\error{D}{f}^{\text{\xmark}\to y}$ is the fraction of samples in~$D$ which~$f$ wrongly assigns to class~$y$.
We define the corresponding error variations as follows:
\begin{align}
\label{eq:variations:source}
v^s(f,f',D,y) &\defeq \error{D}{f}^{y\to \text{\xmark}} - \error{D}{f'}^{y\to \text{\xmark}}\\
 \label{eq:variations:target}
v^t(f,f',D,y) &\defeq \error{D}{f}^{\text{\xmark}\to y} - \error{D}{f'}^{\text{\xmark}\to y}
\end{align}
These variations provide a sort of distance between~$f$ and~$f'$, reflecting the \emph{wrong-prediction gap} of the two models.
Since in FL the global model is improved gradually, round by round, we expect under benign conditions that the wrong-prediction gaps between consecutive models~$f = G^r$ and~$f' = G^{r+1}$ be relatively close. In contrast, a poisoned model can significantly change the error-rate variations.
By comparing error rates of the updated model with those of the previous model, over a \emph{fixed dataset} and \emph{individually for every label}, we can confidently raise a warning whenever the error rates across subsequent rounds differ significantly from those observed for recent, accepted models.

\vspace{1mm}\noindent\textbf{Design detail.}
To measure distances between error-rate variations, we consider the array~$\vec{v}(f,f',D) \defeq [\vec{v}^s,\vec{v}^t]$, where $\vec{v}^s = [v^s(f,f',D,y)]_{y\in Y}$ and~$\vec{v}^t = [v^t(f,f',D,y)]_{y\in Y}$, as a data point in~$\ZZ^{2\size{Y}}$. We then identify each new update as ``suspicious'' based on the relative distance between its \emph{error-variation point} compared to the previous ones.
For this, we rely on the Local Outlier Factor (LOF)~\cite{LOF}, a method to detect outliers in a dataset by inspecting the proximity of a point to its neighbors compared to the neighbors' own proximity. Given a point~$\vec{x}$ and some neighboring points~$N = \{\vec{x_1},\dots,\vec{x_n}\}$, $\LOF_k(\vec{x};N)$ provides a degree of outlier-ness based on the proximity of~$x$ to its~$k$ closest neighbors; in particular, $\LOF_k(\vec{x};N) > 1$ indicates that~$\vec{x}$ is less clustered than the other points, and potentially an outlier.

The detail of our proposal are specified in Algorithm~\ref{algo:model:auditing}.
By the modularity of our design, \emph{any} entity holding labelled data~$D$ could in principle validate a model~$G$ using such strategy, given sufficiently many previously-accepted models (as summarized in the history). This is reflected in our evaluation of~\our{} in Section~\ref{sec:evaluation}, where we consider different configurations letting the server, the clients, and both of them run~\textproc{Validate}.

\begin{algorithm}
	\caption{Instantiation of the validation function}
	\label{algo:model:auditing}
	\begin{algorithmic}[1]
\small
			\Procedure{Validate}{$G,history,D$}
			\State $(\mathcal{G}^{0},\dots,\mathcal{G}^{\ell}) \gets history$
			\For{$i = 1,\dots,\ell$}
			\State $\vec{v}_i \gets \vec{v}(\mathcal{G}^{i-1},\mathcal{G}^{i},D)$
			\Comment{Trusted metric values}
			\EndFor
			\State $\vec{v_{\ell+1}}\gets \vec{v}(\mathcal{G}^\ell,G,D)$
			\Comment{New metric value}
			
			\State $k \gets \lceil\frac{\ell}{2}\rceil$
			\State $h \gets \lceil\frac{\ell*3}{4}\rceil$
			\State $\tau \gets 0$
			\For{$i = h,\dots, \ell$}
		\State $\phi_i \gets \LOF_k(\vec{v}_i; \vec{v}_{i-1},\dots,\vec{v}_{i-h+1})$\\
		\Comment{Trusted LOF}

		\EndFor	
		\State $\tau \gets \texttt{mean}(\phi_h,\dots,\phi_{\ell -1})$%
		\label{line:algo:validation:threshold}
		\Comment{Local threshold}
		\If{$\phi_\ell > \tau$}
			\State $\texttt{vote} \gets 1$ \Comment{Update is suspicious}
		\Else
			\State $\texttt{vote} \gets 0$ \Comment{Update seems OK}
		\EndIf
		\State \textbf{return} $\texttt{vote}$
		\EndProcedure
	\end{algorithmic}
\end{algorithm}

Algorithm~\ref{algo:model:auditing} considers a run of the FL process starting with a global model~$G^0$ initialized by the server.
For~$r\in \sspan{1}{R}$, let $G = G^{r-1}$ denote the global model shared at the beginning of round~$r$, let~$G' = G^r$ be the updated model derived by aggregating the clients' contributions for round~$r$, and let~$history$ denote the latest~$\ell+1$ accepted models.
To validate the current model~$G'$, we calculate the error-variation array~$\vec{v_r} \defeq \vec{v}(G^{r-1},G^{r},D)$.
Intuitively, we aim at inferring the general trend of~$\vec{v}$ from the sequence~$\mathcal{G}^{0},\dots,\mathcal{G}^{\ell}$ of recent models accepted so far, monitoring how this trend changes over the rounds, and make a decision about~$G^{r}$ depending on whether~$\vec{v_r}$ is in line with the previous observations.
Specifically, we compare~$\vec{v_r}$ with the average LOF value of~$\vec{v}$ calculated from the predictions of the~$\ell+1$ latest accepted models, over dataset~$D$.
Here, $\ell$ is a \emph{look-back parameter} determining how many of the older models influence the decision about the current model.
Intuitively, $\ell$ should be sufficiently large to yield a stable reference value for the metric, and at the same time it should be small enough so that ``too old'' models play no role in the decision.%
\footnote{An old model may present significantly higher error variations than the current model, and this may lead to rejecting a genuine model.}
We discuss how to empirically determine an appropriate value for~$\ell$ in Section~\ref{sec:evaluation}.
To summarize, model~$G^r$ is accepted if and only if the current variation~$\vec{v_r}$ is sufficiently close, in the sense of~LOF, to the corresponding values obtained from the history of accepted models.
Namely, we declare~$\vec{v_r}$ as suspicious if it is detected as an outlier compared to the~$k$ closest variations from~$\{\vec{v_1},\dots,\vec{v_\ell}\}$, for $2 \leq k \leq \ell$.
In our implementation, we set $k = \lceil \ell/2 \rceil$, and use as a rejection threshold~$\tau$ the mean of the outlier factors obtained from the last $\lfloor\ell/4\rfloor$ trusted updates.


\section{Implementation \& Evaluation Results}
\label{sec:evaluation}

In this section, we empirically evaluate the effectiveness of~\our{} against backdoor attacks, in the context of image-classification. We conduct several experiments to analyze the detection accuracy of~\our{} under various configurations, varying system's parameters (look-back window size~$\ell$ and quorum threshold~$q$), data splits among server and clients, and the rounds in which poisoning happens (late \textit{vs} early). We also test~\our{} against an adaptive attack.

\subsection{Implementation Setup}

\vspace{1mm}\noindent\textbf{Datasets.}
We consider two image classification tasks, on CIFAR-10 and FEMNIST respectively, and train in both cases a ResNet18 CNN model~\cite{DBLP:journals/corr/HeZRS15}.
The CIFAR-10 dataset~\cite{krizhevsky2009learning} consists of 60,000 colored images with 32x32 pixels and 24-bit color per pixel (3~color channels), of which 50,000 samples are used for training and 10,000 samples are used for testing.
The Federated Extended MNIST (FEMNIST) dataset is an adaptation to the federated setting of the EMNIST dataset~\cite{EMNIST}, a set of handwritten digits derived from the NIST Special Database~19  and converted to a 28x28 pixel image format, so that it matches MNIST dataset structure. The dataset contains 731,668 training samples and 82,587 test samples.

\vspace{1mm}\noindent\textbf{FL setup.}
We run the FL training process with~100 clients for the CIFAR-10 dataset, and with~3550 clients for FEMNIST.
Conforming with previous work~\cite{AISTATS:BagdasaryanVHES20,AISTATS:McMahanMRHA17}, in each round we select 10 contributing clients uniformly at random, and let them train for~2 local epochs with a learning rate of~0.1.
To emulate a non-IID distribution, we assign data to clients according to the Dirichlet distribution with hyper parameter 0.9~\cite{dirichlet}, so that the clients' datasets are unbalanced w.r.t.~the classes.

\vspace{1mm}\noindent\textbf{Implementation.}
We implemented a federated-learning algorithm to solve the aforementioned image-classification tasks, based on publicly available code from~\cite{AISTATS:BagdasaryanVHES20}.
We implemented all algorithms in in Python using the PyTorch framework. We run the experiments on a server with an Intel i5-9600k CPU, equipped with a Gigabyte Geforce RTX 2070 GPU 8GB, and 64GB of RAM. The server runs Ubuntu 18.04 LTS OS.

\vspace{1mm}\noindent\textbf{Attack strategy.}
We test~\our{} against a model-replacement attack~\cite{AISTATS:BagdasaryanVHES20} operated by a single client.
This is not to restrict the attacker's capabilities, as increasing the number of clients submitting poisoned updates can only affect the attack stealthiness against a detector inspecting individual updates.
In the case of CIFAR-10, we instantiate one of the adversarial subtasks considered in~\cite{AISTATS:BagdasaryanVHES20}, causing `cars' with a striped background to be classified as `birds'.
We modify the model-poisoning attack to operate on FEMNIST, causing the backdoored model to misclassify an entire class towards a target class (label-flipping).
We select the source class so that the adversary has most data, to favor the attacker, and the target class uniformly at random among the remaining classes, to avoid any bias.

\vspace{1mm}\noindent\textbf{Defender configurations.}
To showcase the effectiveness of the feedback loop, we consider different configurations of~\our{}, depending on the entities responsible for validating the model: server-only (\layerone), clients-only (\layertwo), and both (\layerthree).
When the feedback loop is in place (\layertwo{} and \our), the server chooses~10 validating clients uniformly at random, and provides them with the current model and the last~$\ell+1$ previously accepted models ($\ell$ is chosen empirically, as we discuss below). Each client runs the~\textproc{Validate} algorithm (cf.~Algorithm~\ref{algo:model:auditing}) locally, before sending their vote to the server. The server decides to accept or reject the new model based on the quorum threshold~$q$, counting in its own vote for the~\our{} configuration.

\subsection{Evaluation Methodology}

\vspace{1mm}\noindent\textbf{System's parameters and data splits.}
We evaluate the impact of lookback window size~$\ell$ and quorum threshold~$q$  on \our{}'s effectiveness, for~$\ell = 10, 20, 30$ and $3 \leq q \leq 9$.
For the configuration where both clients and server validate the model we consider different data splits~C-S\%, reflecting that the clients hold jointly C\% of the overall data and the server holds the remaining~S\%.
We assign shares 90\%-10\%, 95-5\%, and 99-1\% for CIFAR-10, and 99-1\%, 99.5-0.5\%, and 99.9-0.01\% for FEMNIST, so that the ratio between the amount of data at the server versus the amount of data of one client is roughly the same as in CIFAR-10.
Studying how the detection accuracy of~\our{} varies with the splits provides an empirical validation for the effectiveness of the feedback loop, although with little data compared to a real-world deployment.

\vspace{1mm}\noindent\textbf{Poisoning time.}
We assess the effectiveness of~\our{} under two attack scenarios:
(1) the global model~$G$ has already stabilized (accuracy above~90\%, reached after 10,000 clean rounds of FL), and
(2) $G$ is not yet close to convergence and the validation method is enabled in early rounds.
In case (1), we start with a stable model~$G$ and perform~20 subsequent training rounds before injecting three poisoned updates at rounds 30, 35 and 40, respectively (here, $G$ corresponds to `round~1', but we assume an already stable model). We enable the defense after the first 20 rounds in order to build a look-back window of decent size. We terminate the experiment after 50 rounds.
In case (2), we run FL from scratch and start the defense after 500 initial rounds of training. In these early rounds, as the global model is very unstable, we note that a genuine update may cause a significant variation in the per-class error rates and, therefore, it could be mistakenly flagged as malicious (false positive) and discarded.
As a consequence, enabling the defense in early rounds may cause a delay in the convergence of the global model. We therefore consider the very first 800 training rounds and let the attacker inject two malicious updates, at rounds~100 and 300 respectively, before enabling the defense, and subsequently other 10 injections every 15 rounds, starting at round 530.
We deliberately activate the defense after the adversary operated a few injections, to analyze the behavior of~\our{} in case the assumption about early models being trustworthy is violated.
In both cases, we measure the detection accuracy of~\our{} by averaging the results over the 5 repeated experiments as above.

\subsection{Results}

We now report on the results of our experiments to evaluate the effectiveness of~\our{}, in terms of false-negative (FN) and false-positive (FP) rates, under various configurations.

\vspace{1mm}\noindent\textbf{Choice of look-back window size ($\ell$).}
Table~\ref{tab:windowsize} shows the impact of the look-back window size~$\ell$ on FP and FN rates.

\begin{table*}
\centering
\caption{Detection rates of~\layertwo{} (C), \layerone{} (S), and~\our{} (C$+$S) for variable look-back window size~$\ell$ and different data splits C-S\% among clients and server, in CIFAR-10 and FEMNIST.}
\begin{tabular}{|c||c|ccc|ccc||c|ccc|ccc|}
\hline
\multirow{3}{*}{$\ell$} & \multirow{3}{*}{\rotatebox[origin=c]{90}{Split}} & \multicolumn{6}{c||}{CIFAR-10} & \multirow{3}{*}{\rotatebox[origin=c]{90}{Split}} & \multicolumn{6}{c|}{FEMNIST}\\
& & \multicolumn{3}{c}{False Positive Rate} & \multicolumn{3}{c||}{False Negative Rate} & & \multicolumn{3}{c}{False Positive Rate} & \multicolumn{3}{c|}{False Negative Rate}\\
& & C & S & C$+$S & C & S & C$+$S && C & S & C$+$S & C & S & C$+$S\\
\hline
10 &
%
\multirow{3}{*}{\rotatebox[origin=c]{90}{90\%}}&
0.007{\tiny$\pm$0.014} & 0.193{\tiny$\pm$0.053} & 0.021{\tiny$\pm$0.017} & 0.1{\tiny$\pm$0.2} & 0 & 0 &
%
\multirow{3}{*}{\rotatebox[origin=c]{90}{99\%}} &
0 & 0 & 0 & 1 & 1 & 1\\
20 &&%
0.021{\tiny$\pm$0.017} & 0.114{\tiny$\pm$0.042}& 0.029{\tiny$\pm$0.014} & 0 & 0 & 0&
& 0 & 0& 0 & 0& 0& 0\\
30 &&%
0 &0.158{\tiny$\pm$0.094}& 0 & 0.6{\tiny$\pm$0.49}& 0 & 0.1{\tiny$\pm$0.2}&
& 0 & 0.01{\tiny$\pm$0.02}& 0 & 0 & 0 & 0\\
\hline
10 & 
\multirow{3}{*}{\rotatebox[origin=c]{90}{95\%}} &
0.021{\tiny$\pm$0.029} & 0.164{\tiny$\pm$0.029} & 0.021{\tiny$\pm$0.029} & 0.1{\tiny$\pm$0.2} & 0 & 0&
\multirow{3}{*}{\rotatebox[origin=c]{90}{99.5\%}} &
0 & 0 & 0 & 1 & 1 & 1\\
20 &&
0.029{\tiny$\pm$0.027} & 0.193{\tiny$\pm$0.086} & 0.043{\tiny$\pm$0.035} & 0 & 0 & 0 &
& 0 & 0.029{\tiny$\pm$0.027}& 0 & 0 & 0 & 0\\
30 &&
0.021{\tiny$\pm$0.026} & 0.168{\tiny$\pm$0.102}& 0.032{\tiny$\pm$0.042} & 0.4{\tiny$\pm$0.374}& 0 & 0.2{\tiny$\pm$0.245}&
& 0 & 0.12{\tiny$\pm$0.075}& 0 & 0 & 0 & 0\\
\hline
10 & 
\multirow{3}{*}{\rotatebox[origin=c]{90}{99\%}} &
0.014{\tiny$\pm$0.017} & 0.171{\tiny$\pm$0.052} & 0.036{\tiny$\pm$0.032} & 0 & 0 & 0&
%
\multirow{3}{*}{\rotatebox[origin=c]{90}{99.9\%}} &
0 & 0 & 0 & 1& 1& 1\\
20 && 
0.029{\tiny$\pm$0.027} & 0.136{\tiny$\pm$0.042}& 0.029{\tiny$\pm$0.027}& 0 & 0 & 0&
& 0 & 0.179{\tiny$\pm$0.087}& 0 & 0 & 0 & 0\\
30 &&
0.011{\tiny$\pm$0.021} & 0.021{\tiny$\pm$0.042}& 0.011{\tiny$\pm$0.021}& 0.2{\tiny$\pm$0.4}& 0.4{\tiny$\pm$0.49}& 0 &
& 0 &0.22{\tiny$\pm$0.194}& 0 & 0.1{\tiny$\pm$0.2}& 0 & 0.1{\tiny$\pm$0.2}\\
\hline
\end{tabular}
\label{tab:windowsize}
\end{table*}

We study the effectiveness of our proposals for variable~$\ell$ and a default value~$q = 5$.
Concerning CIFAR-10, all configurations using the feedback loop (\layertwo{} and~\our{}) yield good detection rates, i.e., low FN rates between~$0.0$ and~$0.1$, for $\ell = 10,20$, independently of the data split. The~\layertwo{} configuration yields the worst FN rate of~$0.6$ for $\ell = 30$ in the case of~90-10\% split.
Increasing the lookback window~$\ell$ seems to always decrease the number of models detected as poisoned (thus reducing FP and increasing FN). This might be due to a neighborhood that has too many small outliers and therefore reduces the outlier factor of new updates.
In terms of false positives, the feedback loop (\layertwo{} and~\our{}) is clearly superior to the server-only configuration (\layerone), achieving FP rates within~0.0-0.043 \textit{vs}~0.11-0.19 for~$\ell = 10,20$, and~0.0-0.032 \textit{vs}~0.021-0.193 for~$\ell = 30$.
Varying~$\ell$ seems to have a more critical impact in the case of FEMNIST. Indeed, for~$\ell = 10$ none of the configurations is able to detect poisoning attempts, yielding a FN rate of~$1.0$ in all cases. This phenomenon suggests that the look-back window is too small.  Interestingly, all configurations show great improvements as the look-back window is increased to~$\ell = 20,30$, achieving a FN rate of~0 in most cases, and of at most~0.1 in all cases.
Coming to the false positives, again we observe that for~$\ell = 10$ none of the (genuine) models is flagged as positive, as the FP rate is~0 in all cases. As for CIFAR-10, we again observe the feedback loop performing better than the server-only configuration, with FP rates of~0 (for~\layertwo{} and~\our) in contrast to~0.01-0.22 (\layerone).

\begin{figure*}[t!]
	\centering
	\begin{subfigure}[t]{0.322\textwidth}
		\centering
		\includegraphics[width=\linewidth]{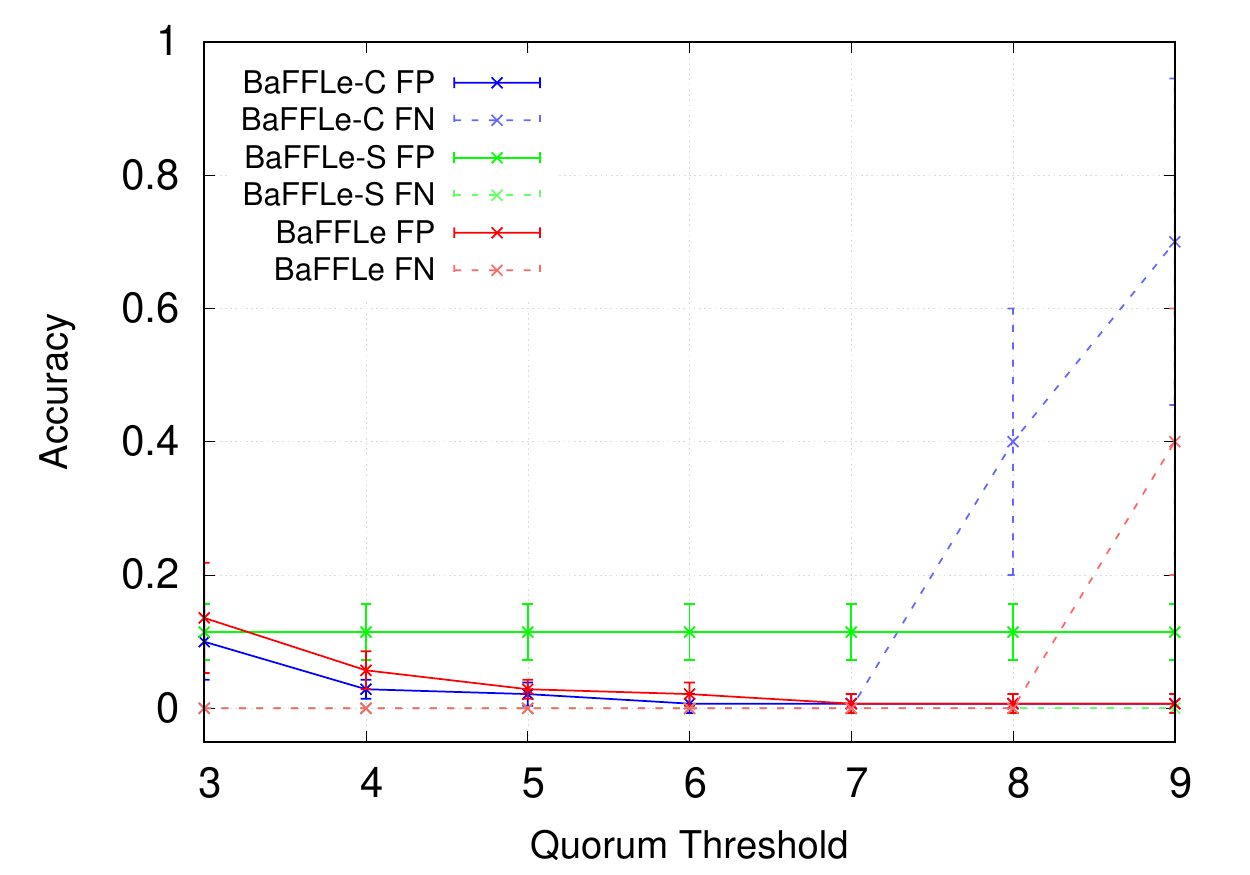}
		\caption{\centering Data split 90\%-10\%, CIFAR}
		\label{fig:quorum:15:cifar}
	\end{subfigure}%
	~
	\begin{subfigure}[t]{0.322\textwidth}
		\centering
		\includegraphics[width=\linewidth]{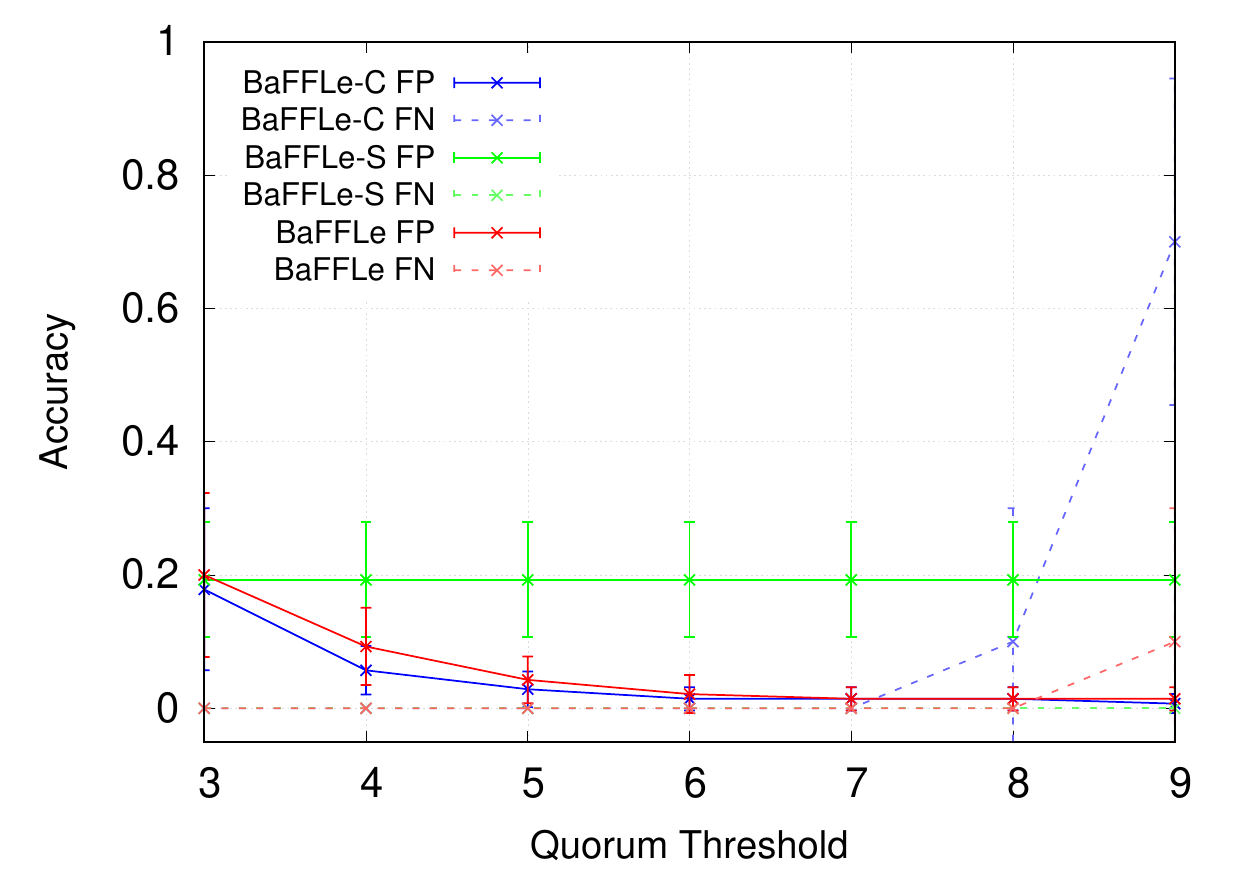}
		\caption{\centering Data split 95\%-5\%, CIFAR}
		\label{fig:quorum:10:cifar}
	\end{subfigure}
	~
	\begin{subfigure}[t]{0.322\textwidth}
		\centering
		\includegraphics[width=\linewidth]{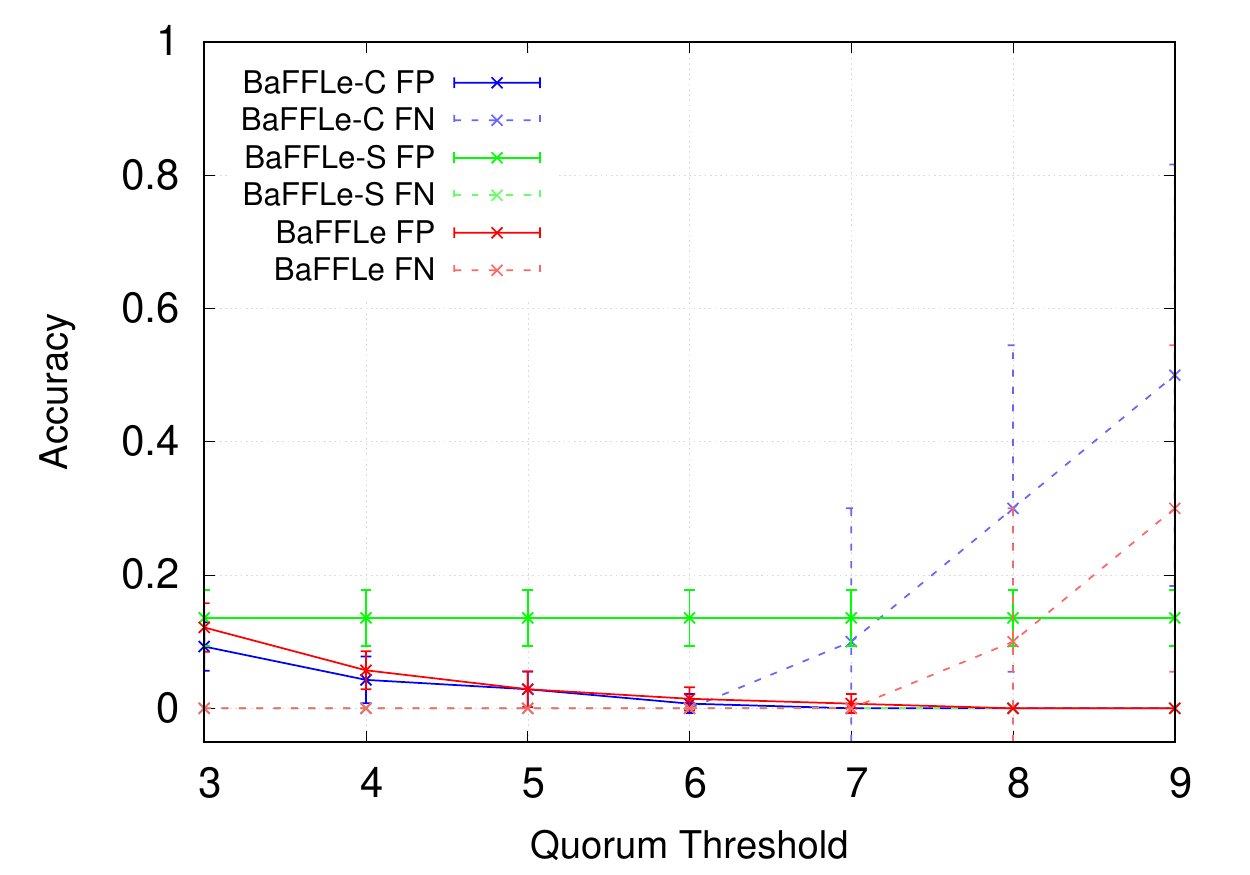}
		\caption{\centering Data split 99\%-1\%, CIFAR}
		\label{fig:quorum:05:cifar}
	\end{subfigure}\\
	\begin{subfigure}[t]{0.322\textwidth}
		\centering
		\includegraphics[width=\linewidth]{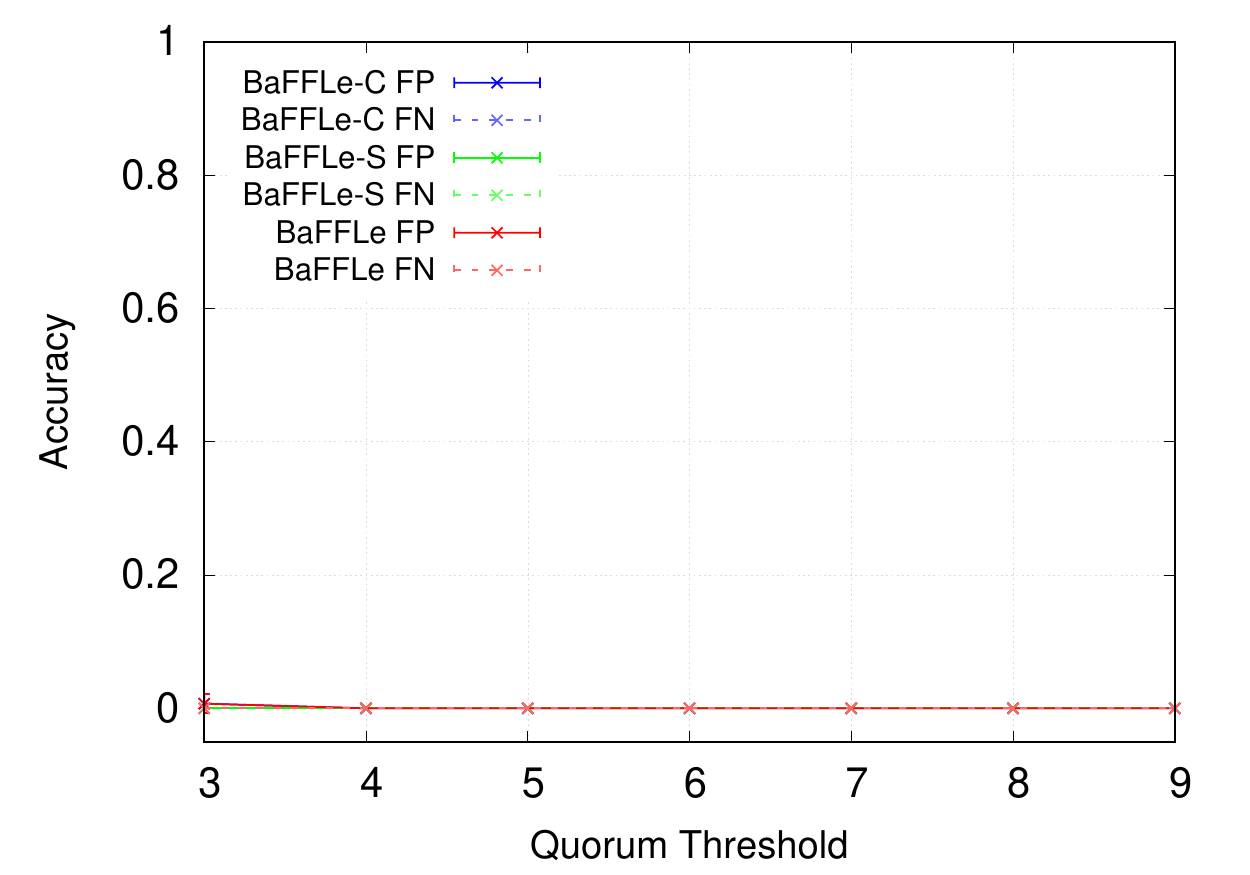}
		\caption{\centering Data split 99\%-1\%, FEMNIST}
		\label{fig:quorum:15:femnist}
	\end{subfigure}%
	~
	\begin{subfigure}[t]{0.322\textwidth}
		\centering
		\includegraphics[width=\linewidth]{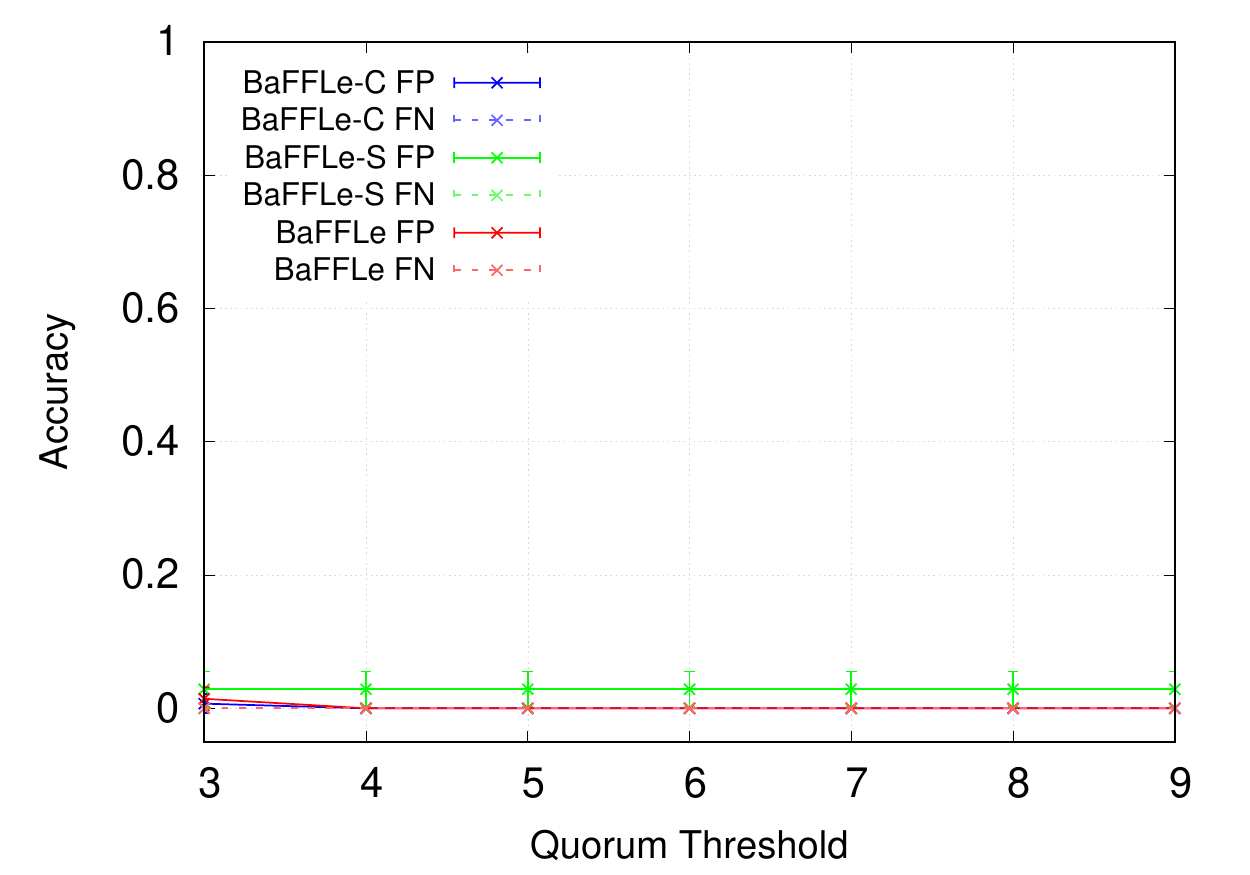}
		\caption{\centering Data split 99.5\%-0.5\%, FEMNIST}
		\label{fig:quorum:10:femnist}
	\end{subfigure}
	~
	\begin{subfigure}[t]{0.322\textwidth}
		\centering
		\includegraphics[width=\linewidth]{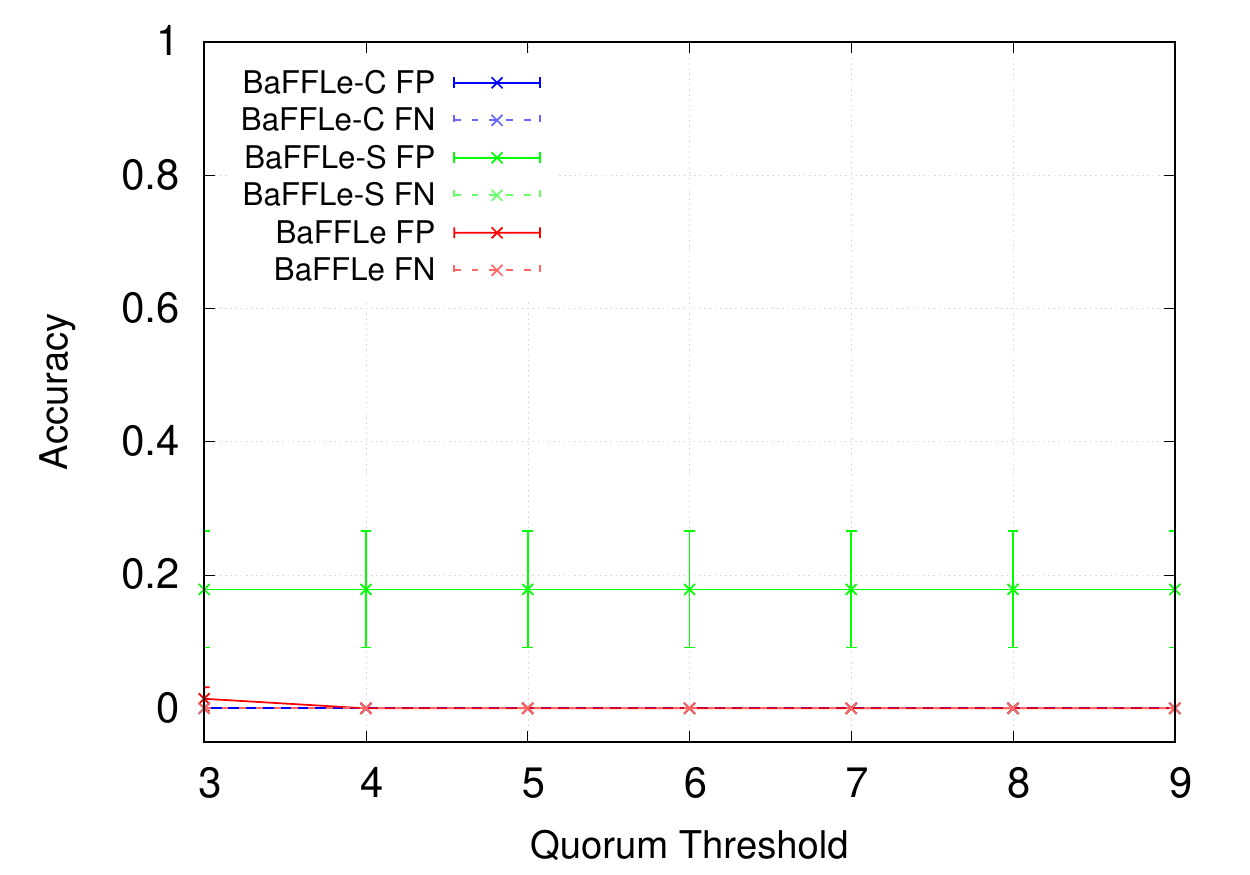}
		\caption{\centering Data split 99.9\%-0.1\%, FEMNIST}
		\label{fig:quorum:05:femnist}
	\end{subfigure}
	\caption{Detection rates (FP and FN) of~\layertwo, \layerone, and~\our{} for variable quorum threshold~$q$ and different data splits among clients and server, in CIFAR-10 {(Figures~\ref{fig:quorum:15:cifar}, \ref{fig:quorum:10:cifar}, and~\ref{fig:quorum:05:cifar}) and FEMNIST (Figures~\ref{fig:quorum:15:femnist}, \ref{fig:quorum:10:femnist}, and~\ref{fig:quorum:05:femnist})}.}
	\label{fig:quorum}
	\vspace{-1 em}
\end{figure*}

 Based on this results, we therefore set~$\ell = 20$ in the sequel, as this value yields the closest to equal error rate for \our.

\setlength\tabcolsep{5pt}
\begin{table}[t]
	\caption{FN rates of~\layertwo{} (C), \layerone{} (S), and~\our{} (C$+$S) against adaptive injections, for different data splits C-S\% among clients and server, in CIFAR-10.}
	\label{tab:adaptive}
	\begin{center}
		\begin{tabular}{|c|l|ccc|}
			\hline
			&\multirow{2}{*}{Attack Type}&\multicolumn{3}{c|}{False Negative Rate}\\
			&&C & S & C$+$S\\
			\hline
			\multirow{2}{*}{\rotatebox[origin=c]{90}{90\%}}
			&Non-Adaptive & 0 & 0 & 0\\
			&Adaptive     & 0.111& 0.333& 0\\
			\hline
			\multirow{2}{*}{\rotatebox[origin=c]{90}{95\%}}
			&Non-Adaptive & 0 & 0 & 0\\
			&Adaptive     & 0 & 0 & 0\\
			\hline
			\multirow{2}{*}{\rotatebox[origin=c]{90}{99\%}}
			&Non-Adaptive & 0.111 & 0 & 0\\
			&Adaptive     & 0 & 0.333 & 0\\
			 \hline
		\end{tabular}
	\end{center}
	\vspace{-1 em}
\end{table}

 \vspace{1mm}\noindent\textbf{Choice of the quorum threshold ($q$).}
The plots in Figure~\ref{fig:quorum} show how the quorum threshold~$q$ affects the detection accuracy of our defense, for both feedback-loop configurations~\layertwo and~\our, under different data splits among clients and server, along with the detection accuracy of the server-only configuration---which is constant as it does not depend on~$q$.
In the case of CIFAR-10 (Figures~\ref{fig:quorum:15:cifar}-\ref{fig:quorum:05:cifar}), varying~$q$ affects significantly the detection accuracy of the feedback loop.
As expected, decreasing~$q$ improves the FN rate, which sharply approaches~0 for~$q \leq 7$ for both~\layertwo{} and~\our, at the expense of slightly increasing the FP rates when~$q$ is decreased further.
We observe that for~$q\leq 5$, \our{} yields slightly higher FP rates compared to~\layertwo, however, it outperforms~\layertwo{} for~$q>6$. Overall, choosing~$5\leq q \leq 7$ appears to be a safe choice as it yields high detection accuracy and nearly equal error rate.
We also note that, for such choice of~$q$, both feedback-loop configurations outperform the server-only configuration, with a FP rate of nearly~0 \textit{vs} about~0.2 for~\layerone.
This behavior is visible for all three data splits.
As for FEMNIST (Figures~\ref{fig:quorum:15:femnist}-\ref{fig:quorum:05:femnist}), changing the quorum threshold does not seem to impact the detection accuracy of~\layertwo{} and~\our{}: all values~$3\leq q\leq 9$ lead to FN and FP rates of~0, regardless of the data split. This behavior is actually not surprising, as all clients detect the attack in this case.

\begin{figure*}[t]
	\centering
	\begin{subfigure}[t]{0.43\textwidth}
		\centering
		\includegraphics[width=0.87\linewidth]{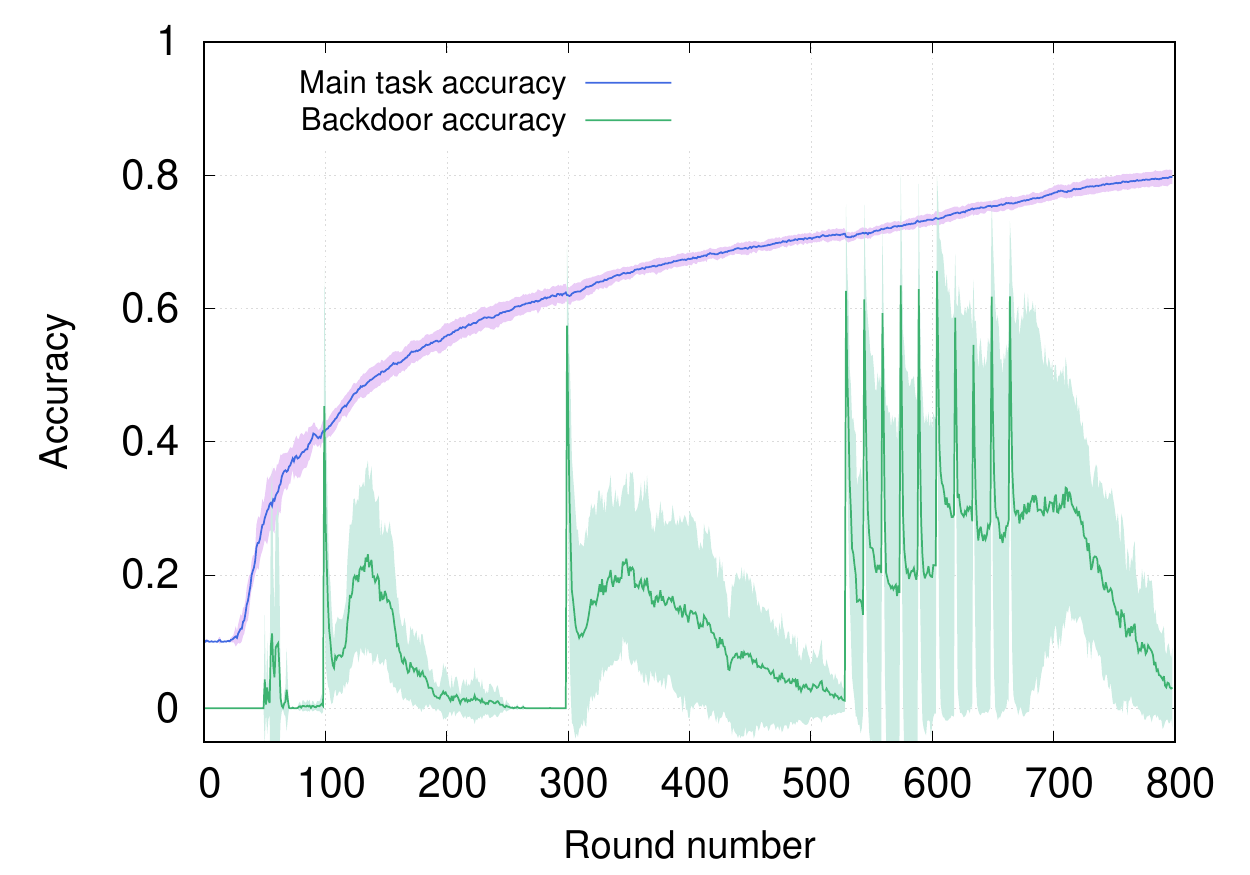}
		\caption{CIFAR without defense}
		\label{fig:poison:cifar:early:wodefense}
	\end{subfigure}%
	~
	\begin{subfigure}[t]{0.43\textwidth}
		\centering
		\includegraphics[width=0.87\linewidth]{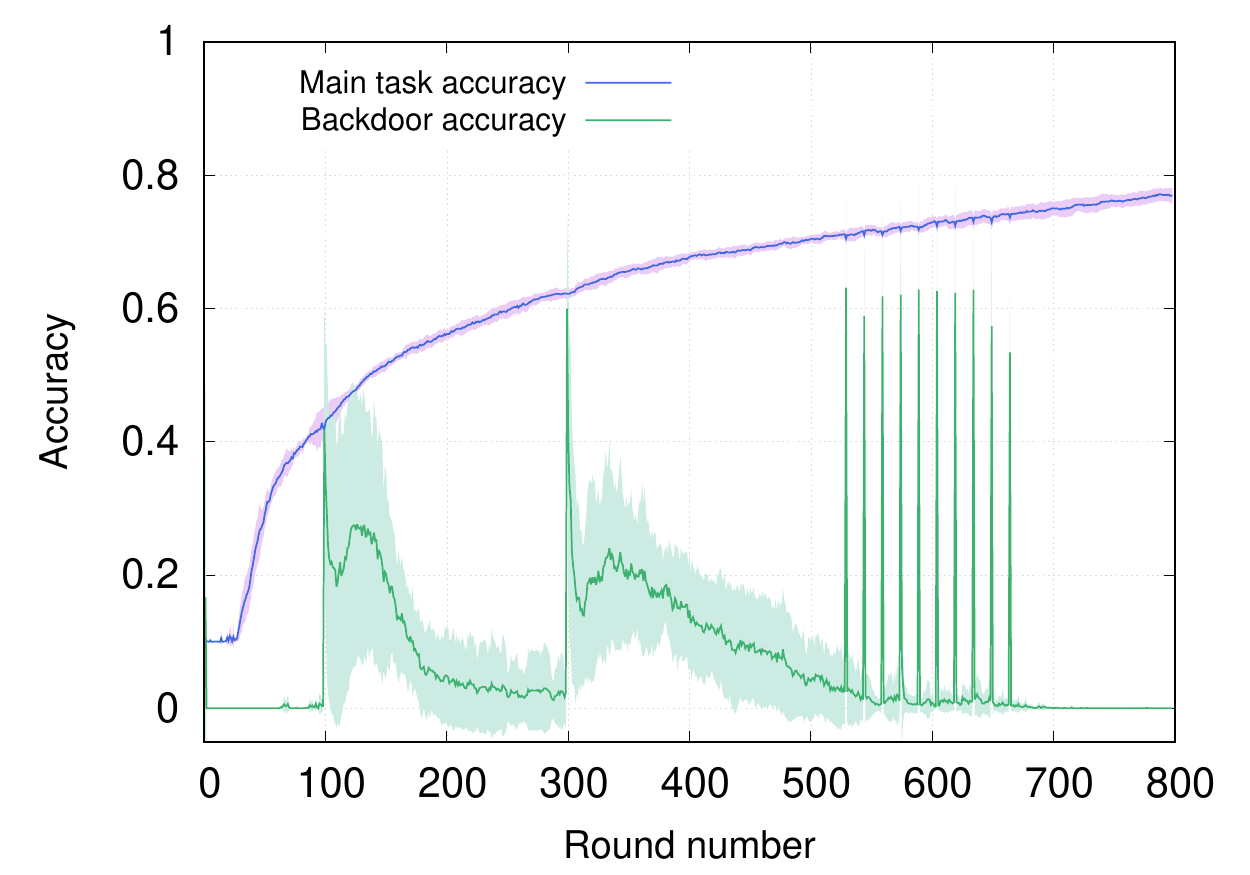}
		\caption{CIFAR with defense}
		\label{fig:poison:cifar:early:wdefense}
	\end{subfigure}%
\\
	\begin{subfigure}[t]{0.43\textwidth}
	\centering
	\includegraphics[width=0.87\linewidth]{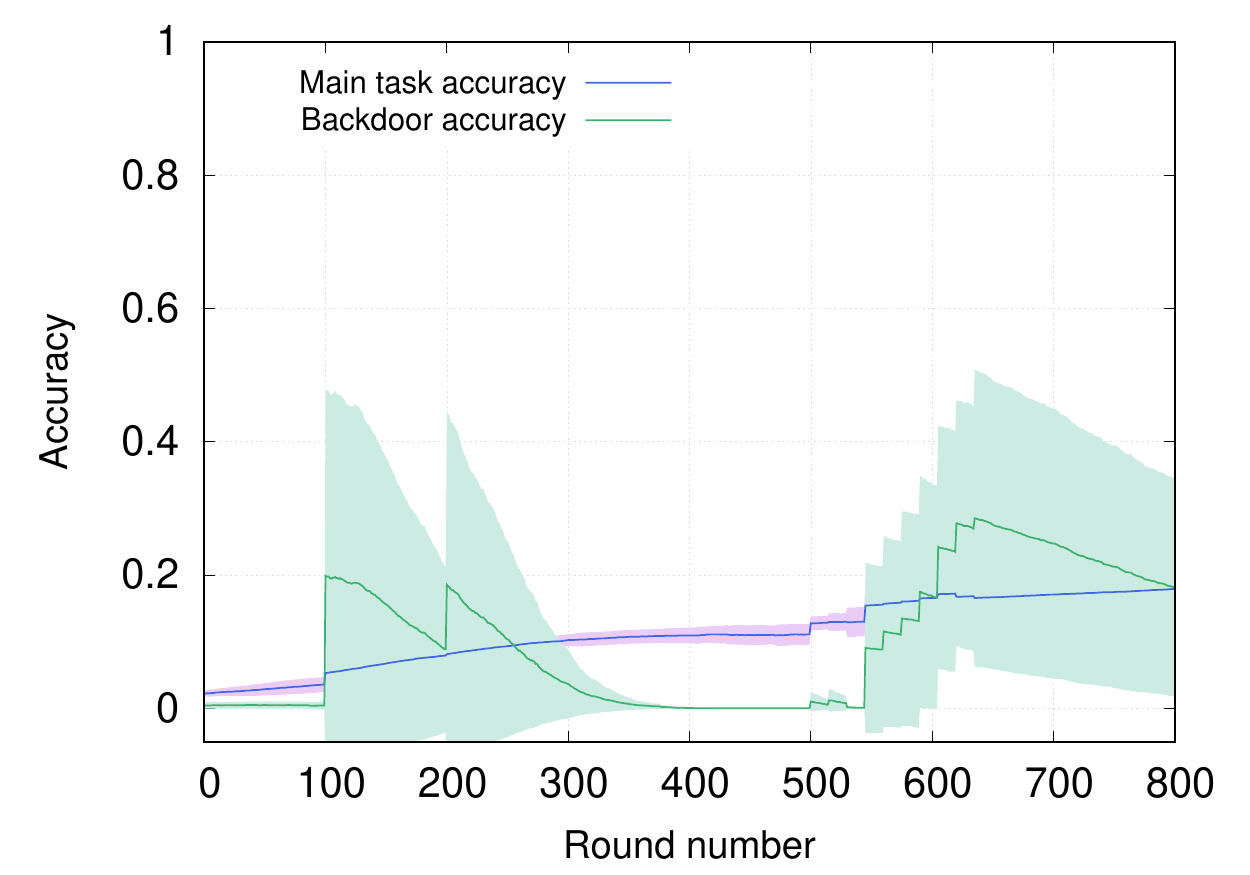}
	\caption{FEMNIST without defense}
	\label{fig:poison:femnist:early:wodefense}
\end{subfigure}%
~
\begin{subfigure}[t]{0.43\textwidth}
	\centering
	\includegraphics[width=0.87\linewidth]{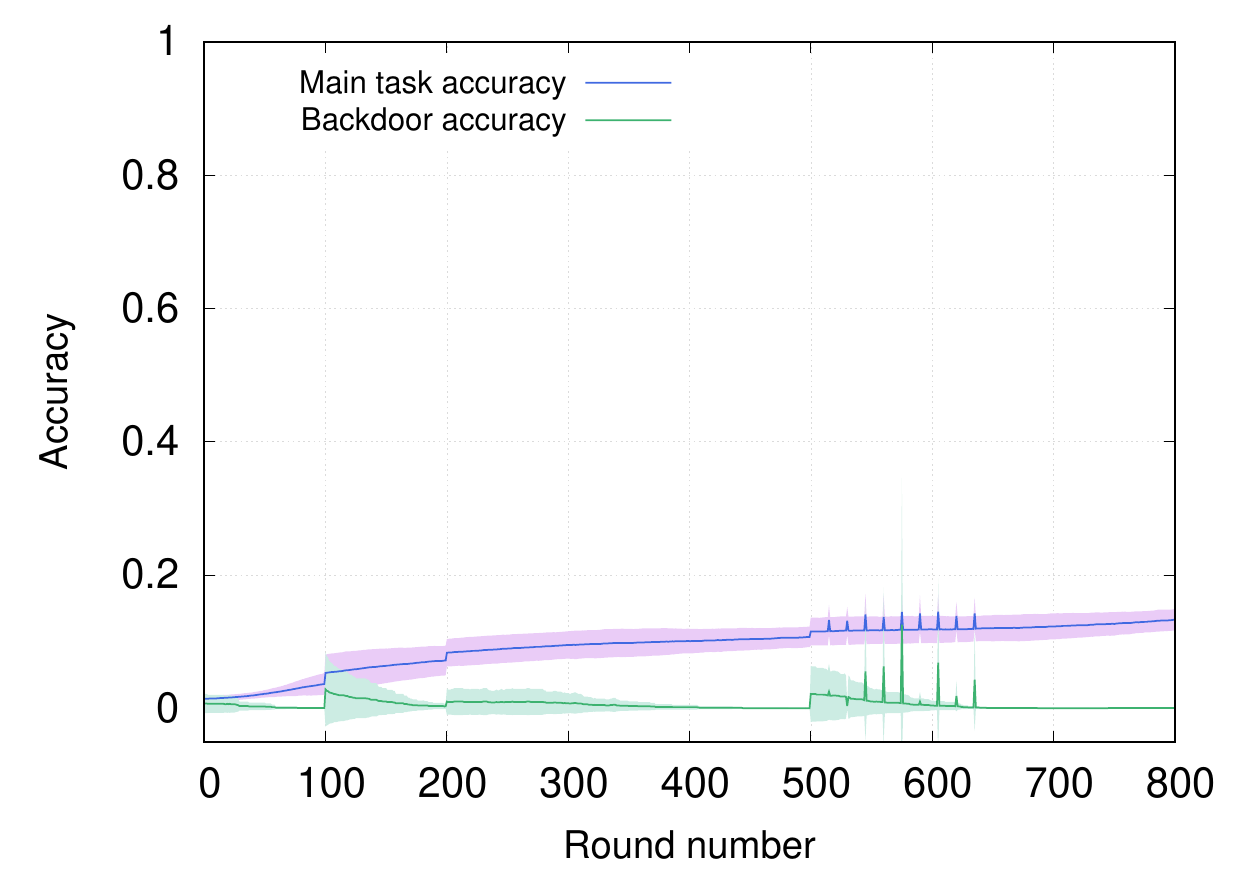}
	\caption{FEMNIST with defense}
	\label{fig:poison:femnist:early:wdefense}
\end{subfigure}%
	\caption{Effect of early poisoning on main-task and backdoor accuracy, without defense (\ref{fig:poison:cifar:early:wodefense} and~\ref{fig:poison:femnist:early:wodefense}) and with~\our{} activated at round 500 (\ref{fig:poison:cifar:early:wdefense} and~\ref{fig:poison:femnist:early:wdefense}). Attacker poisons at rounds~100, 300, 550, then every 15~rounds until~680.}
	\label{fig:poison:early}
	\vspace{-1 em}
\end{figure*}

\vspace{1mm}\noindent\textbf{Detection enabled in early rounds.}
We now analyze the effectiveness of~\our{} in a setting where the model is not yet stabilized, and in the presence of early poisoning.
For this, we consider the very first 800 rounds of training.
We activate the detection method at round~530, when the model starts stabilizing. Nevertheless, we weaken the assumption about early rounds being poison-free and let the adversary inject poisoned updates at rounds~100 and~300 (these injections cannot be detected because the defense is not enabled).
We further let the attacker inject 12 malicious updates at between rounds 530 and 680.
The effect of poisoning on the model's main-task and backdoor accuracy is depicted in Figure~\ref{fig:poison:early}.

In CIFAR-10 (Figures~\ref{fig:poison:cifar:early:wodefense}-\ref{fig:poison:cifar:early:wdefense}), we note that the backdoors injected in early rounds (100 and 300) are not durable, as the backdoor accuracy decreases sharply within a round. This behavior can be justified by the model being so immature that quickly ``forgets'' the backdoor, and is in line with previous studies~\cite{AISTATS:BagdasaryanVHES20,ICML:BhagojiCMC19}. Therefore, these early injections do not affect the model's performance in the long run.
In FEMNIST (Figures~\ref{fig:poison:femnist:early:wodefense}-\ref{fig:poison:cifar:early:wdefense}) we observe a similar behavior, although less prominently due to the slower convergence of the model.

We note that \our{} successfully detects nearly all poisoning attempts operated after round~$530$; only one injection was undetected (in the case of FEMNIST).
Indeed, our detection method conservatively flags new contributions as suspicious as soon as they cause higher variations in the per-class error rates, which is what we leverage to detect model poisoning.

\vspace{1mm}\noindent\textbf{Adaptive attacks.}
We now analyze the effectiveness of our proposal in deterring an adaptive adversary which is aware of the detection method in place and knows the system global parameter~$\ell$ and~$q$.
We focus on the semantic-backdoor attack on CIFAR-10. 
To bypass the validation method based on per-class misclassification analysis, the attacker crafts its updates so that only the backdoor samples in its dataset are misclassified. 
We argue that even such an adaptive strategy is hard to mount in FL: while the attacker can ensure all of its clean data are correctly classified by the backdoored model, it cannot control the model's behavior on the clients' data. Since the local datasets of clients are private and typically very diverse from each other, it is difficult for the attacker to adapt its strategy so that the model behaves in a controlled manner on data the attacker does not know.
Our results are summarized in Table~\ref{tab:adaptive}.

\begin{figure}[ht]
	\centering
		\includegraphics[width=0.77\linewidth]{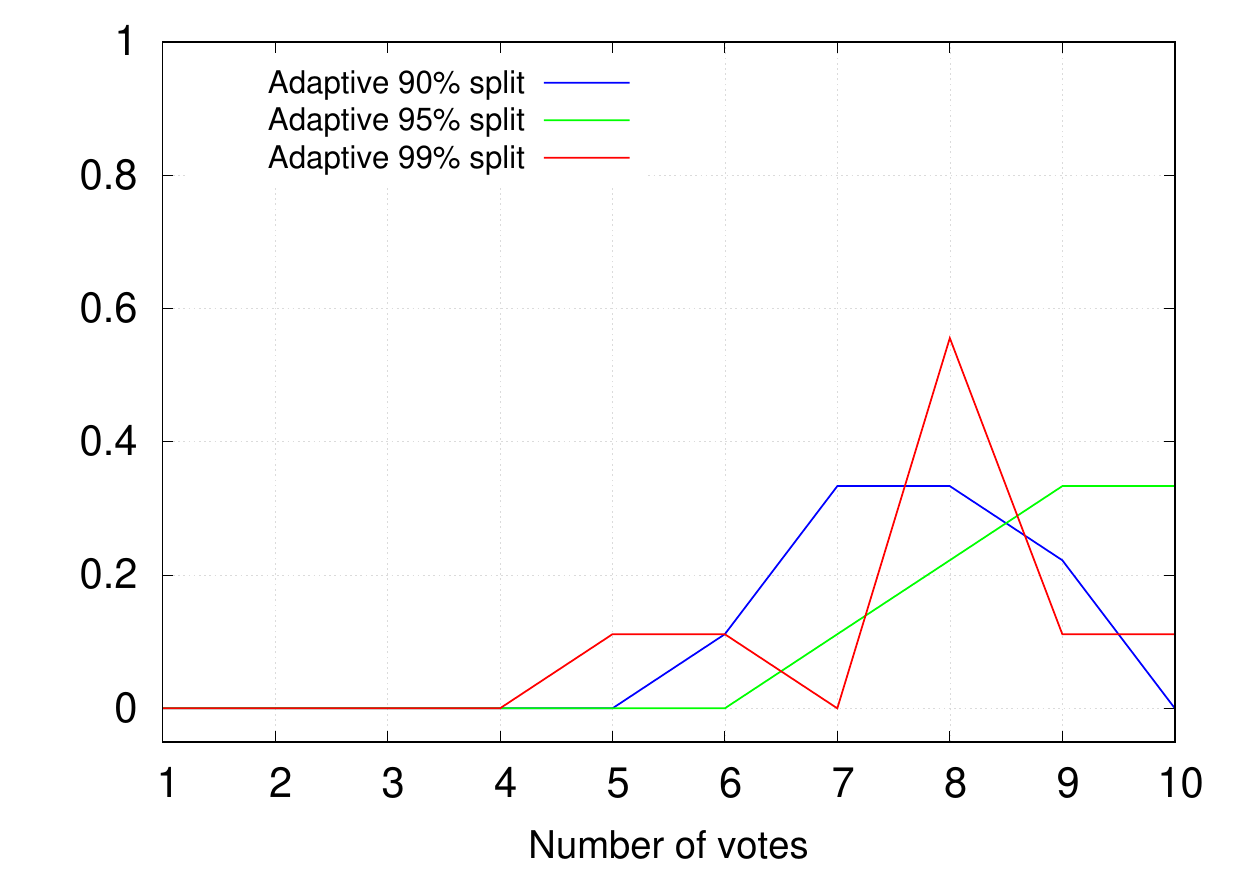}
		\caption{Distribution of votes required to reject adaptively poisoned models, for different data splits among clients and server (90-10\%, 95-5\%, and 99-1\%), in CIFAR-10.}
	\vspace{-1.5 em}
	\label{fig:votes}
\end{figure}

In what follows, we refer to the poisoned injections which remain below the rejection threshold---in the view of the adversary---as \emph{adaptive injections}.
Our results show that between~95\% and~100\% of the adaptive injections are detected by \our, confirming our expectations. Indeed, in FL it is unlikely that the adversary has access to a dataset which is representative of the data of all clients.
Concretely, in CIFAR-10 and for all data splits, \our{} yields~0 false negatives, in contrast to the server-only configuration (\layerone{}) which leads to a 33.3\% FN rate in the 90-10\% and 99-1\% data split. In Figure~\ref{fig:votes}, we further evaluate how different \auditingclients{} perceive the aforementioned adaptive injections. Our results show that most of these injections were detected by~5 or more~\auditingclients.
In other words, at most~5 clients provide a wrong assessment of the model in the worst-case, i.e., $\rho = 0.5$ (cf.~Section~\ref{sec:solution:feedbackloop}). For this value of~$\rho$, we can estimate the threshold on the number of malicious client that we can tolerate as follows: $(1-\rho)(n - n_M) > n_M$, which yields $n_M < (1-\rho) \cdot n/(2-\rho)$. For instance, if $\rho = 0.4$ and $\rho = 0.5$ we have~$n_M < 3.75$, and $n_M < 3.33$, respectively.

\vspace{0.5 em}
\subsection{Communication Overhead}
\label{sec:evaluation:communication}

Our feedback-based defense entails a subset of clients validate each update~$G^r$ for every round~$r$ of training. When combined with the round-based misclassification analysis, it requires validating clients to be equipped with the history of latest~$\ell +1$ accepted models to derive an appropriate rejection threshold (see line~\ref{line:feedbackloop:model:history} in Algorithm~\ref{algo:feedback:loop} and line~\ref{line:algo:validation:threshold} in Algorithm~\ref{algo:model:auditing}).
To save an additional round of communication, we set contributing clients coincide with the validating clients, as in this case the verdict about the global model can be communicated directly, along with the local update.
Namely, in round~$r$ each selected client~$\client_i$ starts with validating model~$G = G^{r-1}$: if the model passes validation, $C_i$ proceeds with training~$G$ to derive a local update~$L_i$, otherwise it rejects the update.
Besides saving one round of interaction between clients and server, this approach has the advantage of letting clients start local training immediately, at the outset of each round, instead of waiting for the verdict from other clients.
In our implementation, the server still needs to send the history to each validating client, which incurs a communication overhead of about~200MB per client that is selected in that round (each model is about~10MB and we set~$\ell = 20$). However, we note that this overhead could be easily reduced by a factor 10, i.e., to around~20MB per selected client, by employing model-compression techniques~\cite{ARXIV:abs-1812-07210}.
Observe that in each round, a client is selected with probability~$\frac{1}{10}$, and hence each client on average is selected to participate and to download~20MB of compressed data about two times every~20 rounds (40MB in total). Notice also that if a client is chosen twice within~20 subsequent rounds, it does not need to download the whole history, as the difference in history from the first time it is selected suffices (i.e., 40MB every~20 rounds is a rather conservative estimate).

\section{Related Work}
\label{sec:related:work}

\vspace{1mm}\noindent\textbf{Attacks.}
A major risk in FL is that malicious clients can manipulate both training data and (local) training algorithm.
Bagdasaryan~\etal~\cite{AISTATS:BagdasaryanVHES20} and Bhagoji~\etal~\cite{ICML:BhagojiCMC19} demonstrate \emph{model poisoning} on FL in the sense of injecting a semantic backdoor into the global model (\emph{targeted} attack) by means of a single, malicious client.
Sun~\etal~\cite{corr:abs-1911-07963} conduct an empirical study of semantic-backdoor attacks on FL, under the assumption that honest clients train with correctly-labelled data which present the backdoor feature, resulting in a strictly weaker setting than that originally considered for model poisoning.
We emphasize that the effectiveness of~\our{} does not rely on honest clients holding (correctly labelled) backdoor data, as we demonstrate in our empirical evaluation (cf.~Section~\ref{sec:evaluation}).
Another poisoning strategy on FL is the \emph{distributed backdoor attack} (DBA) proposed by Xie~\etal~\cite{Xie2020DBA:} in the context of backdoor-trigger attacks. DBA leverages multiple clients to submit poisoned updates containing a  ``trigger portion'' each so that the resulting global model is sensitive to the combined trigger.
In a different vain, Fang~\etal~\cite{fang2019local} present an \emph{untargeted} poisoning attack on FL, aiming at degrading the overall performance of the global model.
This attack shows that various proposals for Byzantine fault-tolerant distributed learning~\cite{NIPS:BlanchardMGS17,ICML:MhamdiGR18,NIPS:BlanchardMGS17,ICML:YinCRB18} are ineffective when applied to the FL setting.

\vspace{1mm}\noindent\textbf{Defenses.}
The first proposal to mitigate poisoning attacks in FL is \emph{FoolsGold} by Fung~\etal~\cite{FoolsGold}, introducing the methodology of inspecting \emph{local} updates and filtering out the suspicious ones. FoolsGold assumes that every class is represented in the data of some honest client, and it relies on the attacker operating through multiple clients. It can be circumvented by an adaptive, single-client attack~\cite{AISTATS:BagdasaryanVHES20}.
Li~\etal~\cite{ARXIV:abs-2002-00211} use spectral anomaly-detection methods to suspect malicious updates in order to defeat both targeted and untargeted attacks.
Nguyen~\etal~\cite{nguyen2021flguard} present~\emph{FLguard}, a two-layer defense to filter out local updates with high backdoor impact and remove residual backdoors via clipping, smoothing, and noise addition. The private version of FLguard guarantees privacy but introduces considerable and costly changes to the FL process.
All these proposals rely on similarity metrics among local updates, which precludes compatibility with secure aggregation.
\our{} instead achieves security while being fully compliant with the standard~FL process---including secure aggregation.
Pillutla~\etal~\cite{pillutla2019robust} presents Robust Federated Aggregation (RFA), lifting the approach of robust distributed learning to the FL scenario. RFA seeks robustness against untargeted attacks degrading the overall classification accuracy; it has been shown vulnerable to targeted attacks~\cite{Xie2020DBA:}.


\section{Concluding Remarks}
\label{sec:conclusion}

Our work tackles the problem of securing FL against backdoor attacks while ensuring compliance with secure aggregation.
We propose~\our, a round-based feedback loop engaging clients in validating the global model.
Our results show that~\our{} can achieve a detection accuracy of~100\% with a false-positive rate below~5\%, on both CIFAR-10 and FEMNIST datasets.
In particular, high detection accuracy is reached even though the validation sets at the clients are relatively small, suggesting that in a real-world deployment, a feedback-based FL system could truly amplify the chances of detecting poisoning attempts.
Moreover, our experiments show that~\our{} is effective even when the detection method is enabled late, as the global model stablizes, and even if there were backdoor injections before, indicating a good level of robustness to early poisoning.
\our{} leverages clients' private data as a source of unpredictability for the defense, making it challenging for an attacker to adapt its strategy for avoiding detection.
Our results confirm this intuition and strongly suggest that crafting adaptive attacks might be harder in the FL setting.
As far as we are aware, \our{} is the first FL defense against backdoor attacks that incurs minimal modifications to existing FL deployments, and efficiently supports privacy of client's data by ensuring full compatibility with secure aggregation of updates. 

\section*{Acknowledgments}

We thank the anonymous reviewers for their valuable feedback.

\bibliographystyle{IEEEtranS}
\bibliography{IEEEabrv,references}

\end{document}